\documentclass[journal,a4paper]{IEEEtran}
\pdfoutput=1

\usepackage[T1]{fontenc}
\usepackage[utf8]{inputenc}

\usepackage{lipsum}

\usepackage{tikz}
\usetikzlibrary{external}
\usetikzlibrary{positioning,calc}
\usetikzlibrary{shapes}
\usetikzlibrary{decorations.markings, spy, fit, patterns, arrows}
\usetikzlibrary{intersections}

\newcommand\encircle[1]{\raisebox{.5pt}{\textcircled{\raisebox{-.7pt}{\scalebox{0.9}{\strut #1}}}}}

\usepackage{mathtools}

\usepackage{pgfplots}
\pgfplotsset{compat=1.18}
\pgfplotsset{plot coordinates/math parser=false}
\pgfplotsset{footnotesize}

\usepgfplotslibrary{polar,groupplots}
\usetikzlibrary{pgfplots.polar}
\newlength\figureheight
\newlength\figurewidth
\usetikzlibrary{decorations.shapes, shapes.geometric}

\PassOptionsToPackage{hyphens}{url}
\usepackage{hyperref}
\hypersetup{
    breaklinks=true,
    pdfauthor={Carsten Andrich, Tobias F. Nowack, Alexander Ihlow, Sebastian Giehl, Maximilian Engelhardt, Gerd Sommerkorn, Andreas Schwind, Willi Hofmann, Christian Bornkessel, Matthias A. Hein, Reiner S. Thomä},
    pdftitle={BIRA: A Spherical Bistatic Radar Reflectivity Measurement System},
    pdfkeywords={Bistatic radar, radar measurements, radar reflectivity, radar cross sections, antenna radiation patterns, integrated sensing and communication, ultra wideband technology, anechoic chambers, digital twin, robotics and automation}
}

\usepackage{amsmath,amssymb}
\usepackage{cite}
\usepackage{comment}
\usepackage{enumitem}
\usepackage{graphicx}
\usepackage{orcidlink}
\usepackage{placeins}
\usepackage{soul}
\usepackage{xurl}
\usepackage{xcolor}

\usepackage{overpic,pict2e}

\usepackage{makecell}
\usepackage{multirow}

\usepackage[acronym]{glossaries-extra}
\setabbreviationstyle[acronym]{long-short}
\glsdisablehyper
\newacronym{ai}{AI}{artificial intelligence}
\newacronym{adc}{ADC}{analog-to-digital converter}
\newacronym{aml}{AML}{antenna measurement laboratory}
\newacronym{api}{API}{application programming interface}
\newacronym[longplural=antennas under test]{aut}{AUT}{antenna under test}
\newacronym{bira}{BIRA}{\textit{Bistatic Radar}}
\newacronym{cmts}{C-MTS}{Celestia Mechatronic Test Systems}
\newacronym{cpcl}{CPCL}{cooperative passive coherent location}
\newacronym{cad}{CAD}{computer aided design}
\newacronym{dac}{DAC}{digital-to-analog converter}
\newacronym{emsl}{EMSL}{European Microwave Signature Laboratory}
\newacronym{ff}{FF}{far field}
\newacronym{icas}{ICAS}{integrated communication and sensing}
\newacronym{if}{IF}{intermediate frequency}
\newacronym{isac}{ISAC}{integrated sensing and communication}
\newacronym{iot}{IoT}{internet of things}
\newacronym{gnb}{gNB}{base station}
\newacronym{gnss}{GNSS}{global navigation satellite system}
\newacronym{gpio}{GPIO}{general purpose input/output}
\newacronym{lamp}{LAMP}{Laboratory of Target Microwave Properties}
\newacronym{lo}{LO}{local oscillator}
\newacronym{los}{LoS}{line of sight}
\newacronym{mec}{MEC}{mobile edge cloud}
\newacronym{mimo}{MIMO}{multiple-input and multiple-output}
\newacronym{ml}{ML}{machine learning}
\newacronym{ms}{MS}{multi-sensor}
\newacronym{mu}{MU}{multi-user}
\newacronym{mvg}{MVG}{Microwave Vision Group}
\newacronym{ncap}{NCAP}{new car assessment program}
\newacronym{nf}{NF}{near field}
\newacronym{ofdm}{OFDM}{orthogonal frequency-division multiplex}
\newacronym{rcs}{RCS}{radar cross section}
\newacronym{rf}{RF}{radio frequency}
\newacronym{ris}{RIS}{reconfigurable intelligent surfaces}
\newacronym{rx}{Rx}{receiver}
\newacronym{scpi}{SCPI}{Standard Commands for Programmable Instruments}
\newacronym{sdr}{SDR}{software-defined radio}
\newacronym{snr}{SNR}{signal-to-noise ratio}
\newacronym{tx}{Tx}{transmitter}
\newacronym[shortplural=UAS]{uas}{UAS}{uncrewed aircraft system}
\newacronym{ue}{UE}{user equipment}
\newacronym{utm}{UTM}{uncrewed aircraft systems traffic management}
\newacronym{vista}{VISTA}{\textit{Virtual Road -- Simulation and Test Area}}
\newacronym{vna}{VNA}{vector network analyzer}
\newacronym{vru}{VRU}{vulnerable road user}
\newacronym{thimo}{ThIMo}{Thuringian Center of Innovation in Mobility}
\newacronym{tui}{TU Ilmenau}{Technische Universität Ilmenau}
\newacronym{hmt}{RF and Microwave Research Group}{RF and Microwave Research Group}

\newcommand{\az}{\varphi}
\newcommand{\el}{\theta}
\newcommand{\azi}{\varphi_\text{ill}}
\newcommand{\eli}{\theta_\text{ill}}
\newcommand{\azo}{\varphi_\text{obs}}
\newcommand{\elo}{\theta_\text{obs}}
\usepackage{pifont}
\newcommand{\yy}{\ding{51}\,}
\newcommand{\nn}{\ding{55}\;}
\newcommand{\spherebig}{30\,cm}
\newcommand{\spheresmall}{10\,cm}

\title{%
    BIRA: A Spherical Bistatic\\
    Radar Reflectivity Measurement System
}

\author{%
    \IEEEauthorblockN{%
        Carsten~Andrich\raisebox{.5ex}{\orcidlink{0000-0002-4795-3517}},
        Tobias~F.~Nowack\raisebox{.5ex}{\orcidlink{0009-0003-5775-9851}},
        Alexander~Ihlow\raisebox{.5ex}{\orcidlink{0000-0002-9714-4881}},
        Sebastian~Giehl\raisebox{.5ex}{\orcidlink{0009-0008-1672-1351}},
        Maximilian~Engelhardt\raisebox{.5ex}{\orcidlink{0009-0002-9440-8615}},\\
        Gerd~Sommerkorn\raisebox{.5ex}{\orcidlink{0009-0003-1111-322X}},
        Andreas~Schwind\raisebox{.5ex}{\orcidlink{0000-0002-7285-7682}},
        Willi~Hofmann\raisebox{.5ex}{\orcidlink{0009-0003-1157-6662}},
        Christian~Bornkessel\raisebox{.5ex}{\orcidlink{0009-0007-0430-7180}},\\
        Matthias~A.~Hein\raisebox{.5ex}{\orcidlink{0000-0002-8751-7760}},~\IEEEmembership{Senior~Member,~IEEE},
        Reiner~S.~Thomä\raisebox{.5ex}{\orcidlink{0000-0002-9254-814X}},~\IEEEmembership{Life~Fellow,~IEEE}
    }
    \thanks{
        The research was funded by the Federal State of Thuringia, Germany, and the European Social Fund (ESF) under grants 2017 FGI 0007 (project ``BiRa''), 2021 FGI 0007 (project ``Kreatör''), and 2023 IZN 0005 (project ``research initiative digital mobility'').
        The research was also funded by the German Federal Ministry of Research, Technology and Space under grant 16KIS2225 (project ``6Gsens'').
        (\textit{Corresponding author: Carsten Andrich.})

        All authors are with the Thuringian Center of Innovation in Mobility, Ilmenau, Germany (e-mail: carsten.andrich@tu-ilmenau.de).

        Carsten Andrich, Alexander Ihlow, Sebastian Giehl, Maximilian Engelhardt, Gerd Sommerkorn, and Reiner S. Thomä are also with the Institute of Information Technology, Technische Universität Ilmenau, Ilmenau, Germany.

        Andreas Schwind, Willi Hofmann, Christian Bornkessel, and Matthias A. Hein are also with the RF and Microwave Research Group, Technische Universität Ilmenau, Ilmenau, Germany.

        To be published in IEEE Transactions on Antennas and Propagation, vol.~nn, no.~n, pp.~nnnn -- nnnn, Month 2026. DOI: 10.1109/TAP.2026.3659989.
    }
}

\makeatletter
\def\ps@IEEEtitlepagestyle{
        \def\@oddfoot{\mycopyrightnotice}
        \def\@evenfoot{}
}
\def\mycopyrightnotice{
    \begin{minipage}{\textwidth}
        \centering
        \scriptsize
        Copyright~\copyright~2026 IEEE.
        Personal use of this material is permitted.
        Permission from IEEE must be obtained for all other uses, in any current or future media, including reprinting/republishing this material for advertising or promotional purposes, creating new collective works, for resale or redistribution to servers or lists, or reuse of any copyrighted component of this work in other works.
    \end{minipage}
}

\begin{document}

\hbadness=10000

\maketitle

\begin{abstract}
The upcoming 6G mobile communication standard will offer a revolutionary new feature:
Integrated sensing and communication (ISAC) reuses mobile communication signals to realize multi-static radar for various applications including localization.
Consequently, applied ISAC propagation research necessitates to evolve from classical monostatic radar cross section (RCS) measurement of static targets on to bistatic radar reflectivity characterization of dynamic objects.
Here, we introduce our Bistatic Radar (BIRA) measurement facility for independent spherical positioning of two probes with sub-millimeter accuracy on a diameter of up to 7\,m and with almost continuous frequency coverage from 0.7 up to 260\,GHz.
Currently, BIRA is the only bistatic measurement facility capable of unrestricted ISAC research:
In addition to vector network analysis, it employs advanced wideband transceiver technology with an instantaneous bandwidth of up to 4\,GHz.
These transceivers grant BIRA the unique capability to characterize dynamic targets in both Doppler and range, while also significantly accelerating measurements on static objects.
Additionally, the installation is capable of spherical near-field antenna measurements over these wide frequency ranges.
\end{abstract}


\section{Introduction}
\label{sec:introduction}
\Gls{isac} is considered one of the key features of future 6G mobile communication~\cite{liu24_book_isac}.
In essence, \gls{isac} is a means of radar detection and localization of passive objects, i.e., radar targets, that are not equipped with a radio tag.
The radio signals transmitted for communication purposes by base stations or mobile radio nodes are reused for target illumination.
\Gls{isac} enables to gain comprehensive situational awareness and to improve efficiency and safety in road traffic scenarios, such as V2X (vehicle-to-vehicle or vehicle-to-infrastructure) and for the monitoring of the lower altitude air space (U-space).
Moreover, there are many other potential use cases of \gls{isac} for public security and safety applications and for increased efficiency in mobile radio access~\cite{shatov24_access_jrc}.

Whereas most conventional, e.g., automotive, radar systems are monostatic, the \gls{isac} sensing geometry is bistatic, because it corresponds to the spatial separation of user equipment and infrastructure in mobile communications.
Therefore, multi-sensor \gls{isac}~\cite{thomae19_commag_cpcl, thomae23_eurad_distributed_isac, thomae23_distrib_multi_isac} is generally equivalent to a distributed \gls{mimo} radar~\cite{haimovich08_spmag_mimo_radar}.
A related advantage over conventional monostatic radar is the inherent diversity gain, which increases the probability of target detection and supports target recognition.

Based on dynamic ray tracing and multilink wireless channel sounders~\cite{thomae23_arxiv_multilink_isac}, advanced simulation and measurement techniques already enable the performance evaluation of distributed ISAC systems.
However, the investigation of wave interaction at solitary targets in multistatic scenarios still deserves special attention in propagation research.
As all multistatic scenarios are superpositions of bistatic ones, this interaction is best described by the target-specific, complex-valued, bistatic reflectivity function, which comprises the following dimensions:
\begin{itemize}
    \item[(i)] The 4-dimensional constellation of the illuminating \gls{tx} and the observing \gls{rx} in spherical coordinates around the target with azimuth angles~$\azi$ and $\azo$ as well as co-elevation angles~$\eli$ and $\elo$ for illumination and observation, respectively.
    \item[(ii)] The 2-dimensional polarizations at \gls{tx} and \gls{rx}.
    \item[(iii)] The frequency response, which corresponds to the bistatic time of flight, i.e., the bistatic depth of the target.
    \item[(iv)] The temporal variation of the target reflectivity due to its internal, locally confined movements, which corresponds to micro-Doppler~\cite{chen19_book_udoppler_radar}.
\end{itemize}

The reflectivity function describes the respective target exclusively and comprises no other objects nor superimposed propagation effects.
Meticulous measurement and post-processing techniques are required to de-embed all such effects, e.\,g., radiation patterns of antennas in their installed state, multipath propagation due to parasitic reflections, path loss, and the frequency response of the entire measurement setup.
Appropriate measures include the extensive use of microwave absorbers, antenna and system calibration~\cite{andrich25_eucap_antenna_deconvolution_radar_refl}, near-field to far-field transformation, or synthetic aperture processing, which are beyond the scope of this paper.
Knowledge of de-embedded, i.\,e., undistorted, reflectivity functions of various targets enables their universal use within simulations as well as realistic 6G \gls{isac} object emulation.
The research objective is to combine measured target reflectivities with channel models~\cite{ebrahimizadeh24_tap_rcs_channel_model,liu24_tvt_channel_model_isac} or airborne~\cite{beuster25_wsa_testbed_airborne,beuster25_wsa_isac_drone_data_set} and terrestrial~\cite{stanko23_vtc_multilink_channel_sounding,fenollosa25_globecom_bistatic_isac_channels} propagation measurements to ascertain accurate propagation conditions for a multitude of scenarios.

\begin{figure}
    \centering
    \includegraphics[width=\linewidth]{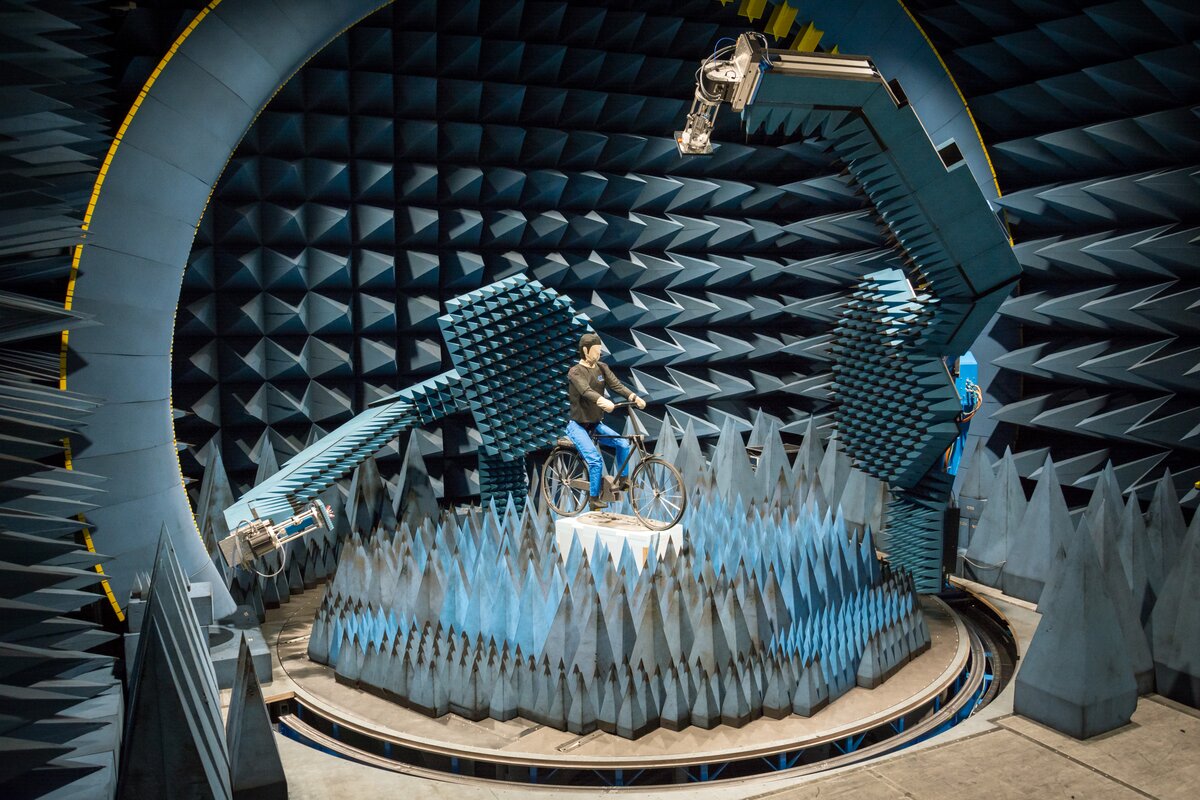}
    \caption{
        The \glsfirst{bira} measurement system installed in the \glsfirst{vista} at the \glsfirst{thimo}.
        A motorized \gls{vru} NCAP dummy is placed on the turntable as an exemplary target for bistatic reflectivity measurement with optional motion for micro-Doppler.
    }
    \label{fig:mvg_bira_raytraced_vru}
\end{figure}

This paper introduces our novel \gls{bira} spherical measurement facility designed to determine the eight-dimensional, bistatic reflectivity function of extended, time-variant targets.
Key components include a programmable mechanical system for independent positioning of \gls{tx} and \gls{rx} antennas with their frontends around the target within an anechoic chamber, accommodating objects as large as a passenger car.
The system enables synchronous signal distribution, high positioning accuracy for spatial coherency, and fully polarimetric measurement.
Optional wideband transceivers feature high instantaneous bandwidth, a short duty cycle, and good correlation properties, enabling high repetition rates to capture target time variability~\cite{thomae00_tim_channel_sounding}, such as micro-Doppler~\cite{chen19_book_udoppler_radar}.
Arbitrary, dynamic waveforms allow for versatile applications beyond de-embedded reflectivity measurement, e.\,g., over-the-air testing with communication signals, multipath channel emulation, or interference suppression tests.
The system is designed to match the frequency range and bandwidth of current and future mobile wireless access schemes, facilitating target shape identification and mitigation of parasitic reflections.

6G will offer signal bandwidths in excess of several hundred MHz.
This enables \gls{isac} with a high spatial resolution in the order of decimeters or better, which is sufficient to locate, resolve, and subsequently classify radar targets.
However, the high resolution is challenged by artifacts originating from \gls{rx} and \gls{tx} antennas as well as their surroundings that superimpose the target response.
Therefore, the characterization, calibration, and compensation of antennas in their installed state~\cite{berlt20_eucap_phase_recovery} presents an inherent aspect of \gls{isac} research, for which \gls{bira} is also well suited.

\Gls{bira}, shown in \autoref{fig:mvg_bira_raytraced_vru}, is a comprehensive hardware upgrade to our existing \gls{vista}.
Over the last decade, VISTA has enabled diverse and novel contributions primarily in the field of automotive wireless applications, e.g., vehicle-in-the-loop virtual drive testing, performance evaluation of automotive antennas in their installed state, radar testing, and realistic emulation of \gls{gnss} signals~\cite{hein23_roe_vista}.
Predating \gls{bira}, we have also utilized \gls{vista} for static~\cite{schwind_2018_rcs_vulnerable_road_user} and dynamic~\cite{schwind20_iet_bistatic_delay_doppler} \gls{isac} measurements, but for lack of suitable mechanical positioning infrastructure, the respective bistatic geometries were essentially restricted to selected horizontal planes across a cylindrical sampling surface.
We also developed a small-scale bistatic measurement system with four degrees of freedom~\cite{eumw17roeding_biarc} and carried out bistatic millimeterwave \gls{rcs} measurements of \glspl{uas}~\cite{roeding17_eucap_rcs_drones}.
The mechanical constraints of our prior work have motivated the \gls{bira} upgrade and have shaped its conception and implementation.

This paper is structured as follows:
\autoref{sec:sota} provides a comparison with state-of-the-art bistatic measurement facilities.
\autoref{sec:mechanics} describes the mechanical aspects of the \gls{bira} system and \autoref{sec:rf} explains its measurement transceivers and interchangeable microwave probe modules.
Then, \autoref{sec:software} illustrates the associated software suite comprising a digital twin and a generic \gls{api} and abstraction layer.
Finally, \autoref{sec:examples} details five measurement examples including results and \autoref{sec:conclusions} summarizes this paper.

\section{State of the Art}
\label{sec:sota}

\begin{table*}
\centering
\caption{Comparison of BIRA and state-of-the-art bistatic measurement facilities from literature.}
\label{table:sota}
\scriptsize
\renewcommand{\arraystretch}{1.1}
\setlength{\tabcolsep}{2pt}
\begin{tabular}{ |p{1.6cm}|p{2.9cm}|p{2.5cm}|p{2.5cm}|p{2.5cm}|p{2.5cm}|p{2.5cm}|  }
\hline
    & \multicolumn{1}{c|}{\textbf{presented here}} & \multicolumn{5}{c|}{\textbf{bistatic measurement facilities described in literature}} \\ \hline
    &	\textbf{BIRA @TU Ilmenau, DE}	&	BIANCHA @INTA, ES	&	CACTUS @CESTA, FR	&	BABI @ONERA, FR	&	EMSL @Ispra, IT	&	LAMP @DASA, CN \\ \hline \hline
 inauguration & 2023 & 2010 & <2007 & 1990s & 1992 & 2019 \\ \hline
 principle & two gantry arm positioners & two elevated scanning arms & two slides on one shared circular azimuth rail & two slides on one shared circular azimuth rail & two slides on one shared elevation arc orbit & two slides on one shared elevation arc orbit \\ \hline
 target distance & $\leq$3.5\,m; w/ RF probe\;$\approx$\;3\,m & 1.74\,m & a few m & 5.5\,m & 10\,m & 9.3\,m \\ \hline
 \# mech.\,DoF$^*$ & 8 & 4 & 2 & 2 & 3 & 3 \\ \hline
 \quad target $\circlearrowleft$ & \yy turntable 360° $\varnothing$ 6.5\,m & \yy turntable 360° $\varnothing$\,<1\,m & \nn pillar & \nn mast, variable height & \yy turntable 360° on rails & \yy turntable 360° on rails \\ \hline
 \quad Tx azimuth & \yy --118°...\,66° & \nn (fixed) & \yy 0°...\,180° & \yy 0°...\,190°& \nn (fixed) & \nn (fixed) \\ \hline
 \quad Rx azimuth & \nn (fixed) & \yy --200°...\,200° & \yy 0°...\,180° & \yy 0°...\,190°& \nn (fixed) & \nn (fixed) \\ \hline
 \quad Tx co-elev. & \yy --114°...\,114° & \yy --100°...\,100° & \nn (fixed, 90°) & \nn (manually, 70°...\,90°) & \yy --115°...\,115° & \yy --90°...\,90° \\ \hline
 \quad Rx co-elev. & \yy --115°...\,115° & \yy --100°...\,100° & \nn (fixed, 90°) & \nn (manually, 70°...\,90°) & \yy --115°...\,115° & \yy --90°...\,90° \\ \hline
 \quad Tx polariz. & \yy --10°...\,188°& \nn & \nn  & \nn  & \nn & \nn \\ \hline
 \quad Rx polariz. & \yy --10°...\,188°& \nn & \nn  & \nn  & \nn & \nn \\ \hline
 \quad Tx radial & \yy 3.44\,m\;$\pm$\;6\,cm & \nn & \nn  & \nn  & \nn & \nn \\ \hline
 \quad Rx radial & \yy 3.44\,m\;$\pm$\;6\,cm & \nn & \nn  & \nn  & \nn & \nn \\ \hline
 bistatic angles$^\ddag$ & more than hemisphere & more than hemisphere & partial azimuth & partial azimuth & partial hemisphere & partial hemisphere \\ \hline
 \quad $\azi$ & 0°...\,360° & 0°...\,360° & 0°...\,180° & 0°...\,190° & 0°...\,360° & 0°...\,360° \\ \hline
 \quad $\azo$ & 0°...\,360° & 0°...\,360° & few°...\,180° & 4°...\,190° & 0° \textbf{or} 180° & 0° \textbf{or} 180° \\ \hline
 \quad $\eli$ & 0°...\,114° & 0°...\,100° & 90° (fixed) & 70°...\,90° (manually) & 0°...\,115° & 0°...\,90° \\ \hline
 \quad $\elo$ & 0°...\,115° & 0°...\,100° & 90° (fixed) & 70°...\,90° (manually) & 0°...\,115° & 0°...\,90° \\ \hline
 measurement instrumentation & VNA (for static DUTs); \newline wideband SDR transceivers \newline (resolve Doppler dynamics) & VNA & VNA & VNA & VNA & VNA \\ \hline
 frequency range & 0.7\,...\,20\,GHz, \newline with conversion: $\leq$260\,GHz & 0.8\,...\,40\,GHz & 1\,...\,18\,GHz & 0.5\,...\,40\,GHz & 0.7\,...\,50\,GHz & 0.8\,...\,20\,GHz \\ \hline
 probes & arbitrary, $\approx\,$50$\times$30$\times$30\,cm, $\leq$20\,kg (antennas, frequency converters, transceivers, sensors, e.g., Lidar, ToF, ...) & antennas & antennas & antennas & antennas & antennas \\ \hline
 additional \newline payload on \newline positioners & mounting plate 38$\times$17\,cm for $\leq$35\,kg payload \newline (e.g., transceivers) &  &  &  &  &  \\ \hline
 reference & [this paper] & \cite{sota_biancha_2010,sota_biancha_2016,biancha_www} & \cite{sota_cactus} & \cite{sota_babi} & \cite{sota_emsl_sieber_1992,emsl_www} & \cite{sota_lamp_tian_2021,lamp_www} \\ \hline
\end{tabular}
\newline \raggedright
\newline $^*$ Number of mechanical degrees of freedom.
Mechanical parameters are given within the respective machine coordinate system.
It differs from the bistatic angles, which are relative to the target.
0° co-elevation locates the probe in the zenith.
\newline $^\ddag$ $\azi, \eli$: Illumination azimuth and co-elevation angle; $\azo, \elo$: observation angles.
All coordinates are relative to the target, which is located at the origin of the spherical coordinate system.
The x-y-plane is aligned with ground and the z-axis points towards the zenith.
Co-elevation angle $\el$ is defined down from the zenith.
Note that the set of reachable azimuth angles $\azi, \azo$ may also incorporate a target rotation with the turntable.
\end{table*}

Only a few measurement facilities are presently known to enable measurements of the bistatic radar reflectivity of extended targets.
Those found in the literature will be introduced in the following.
To facilitate a comparison by features, their capabilities are contrasted with those of \gls{bira} in \autoref{table:sota}.
The \gls{tx} and \gls{rx} positioner characteristics (azimuth and co-elevation) are given with reference to the respective machine coordinate system.
Together with the target rotation enabled by a turntable, the mechanical degrees of freedom of the positioners result in a set of reachable bistatic angles from the target perspective.

1.) The bistatic anechoic chamber (BIANCHA) of the \textit{National Institute for Aerospace and Technology} (INTA) in Torrejón de Ardoz, Spain, consists of a turntable and two gantry arms, providing four degrees of freedom~\cite{sota_biancha_2010,sota_biancha_2016}.

2.) In the CACTUS measurement facility at the \textit{centre d'études scientifiques et techniques d'Aquitaine} (Cesta) in France, two pedestals with elevated antennas can move along a circular rail (on the floor), providing two degrees of freedom~\cite{sota_cactus}.

3.) Similarly, in the BABI measurement chamber of the \textit{office national d'études et de recherches aérospatiales} (Onera) in France, the antennas slide on an elevated circular rail~\cite{sota_babi}.

4.) The \gls{emsl}, located in Ispra, Italy, contains one elevation arch with two sliding antennas and a 360\textdegree{} turntable positioner~\cite{sota_emsl_sieber_1992,emsl_www}.

5.) The \gls{lamp}, located in Deqing, Zhejiang, China, contains two sliding antennas on an elevation arch and the target placement onto a turntable on rails is comparable to \gls{emsl}~\cite{sota_lamp_tian_2021,lamp_www}.

Only BIANCHA and \gls{bira} enable unrestricted, bistatic, spherical positioning, i.e., setting illumination azimuth angle~$\azi$ and co-elevation angle~$\eli$ independently of observation azimuth angle~$\azo$ and co-elevation angle~$\elo$, which we consider mandatory for bistatic radar reflectivity and antenna measurement research.
Exclusively \gls{bira} is large enough for target objects up to the size of a passenger car.
Regarding the measurement instrumentation, all other facilities only use \glspl{vna}.
In contrast, \gls{bira} also features \gls{sdr} transceivers for Doppler- and range-resolved measurement of time-variant targets, which are integral to \gls{isac} research.
Additionally, \gls{bira} supports interchangeable, arbitrary positioner probes and payloads that enable 6G-ready frequency coverage up to 260\,GHz through integrated converters.

Owing to its unique capabilities, \gls{bira} is currently the only published bistatic measurement facility capable of performing unrestricted \gls{isac} research.

\section{Concept and Mechanical Properties}
\label{sec:mechanics}

\Gls{bira} was conceptualized as an upgrade to the \gls{vista} facility depicted in \autoref{fig:mvg_bira_raytraced_vru}, which provides the ideal prerequisites for bistatic reflectivity and antenna measurements of target objects up to the size of a passenger car.
\Gls{vista} comprises a shielded, anechoic chamber with 13\,m~$\times$ 9\,m~$\times$ 7.5\,m of usable interior volume, a turntable with 6.5\,m diameter for loads up to 3000\,kg, and a permanently installed SG~3000F multi-probe antenna measurement arch for frequencies from~70\,MHz up to~6\,GHz~\cite{vista}.

The immovable measurement arch necessitates two significant constraints for the installation of our bistatic positioning system \gls{bira}:
It must fit within the inner diameter of the arch and, more importantly, it must be removable and re-installable to enable unobstructed use of the multi-probe system.
Assembly duration should be no more than a single day and repeated re-installation must not degrade the mechanical accuracy.
The contractor commissioned with the development and installation of \gls{bira} was \gls{cmts}.

\begin{figure}
    \centering
    \begin{tikzpicture}
    \node[anchor=south west,inner sep=0] (image) at (0,0) {\includegraphics[width=1\linewidth]{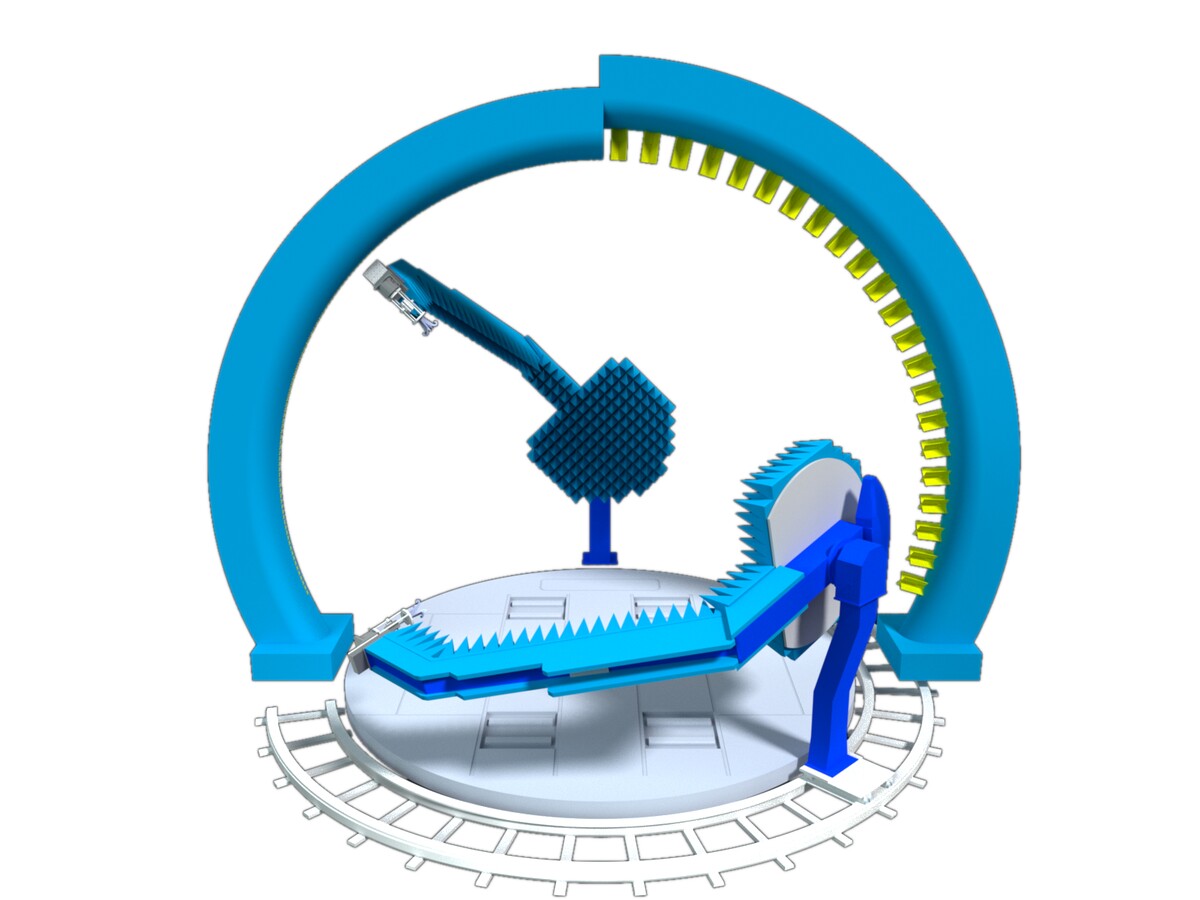}};
    \begin{scope}[x={(image.south east)},y={(image.north west)}]
        \node[minimum width=1ex,inner sep=.25ex,circle,fill=white,draw] () at (0.56,0.59) {a};
        \node[minimum width=1ex,inner sep=.25ex,circle,fill=white,draw] () at (0.32,0.62) {b};
        \node[minimum width=1ex,inner sep=.25ex,circle,fill=white,draw] () at (0.50,0.90) {c};
        \node[minimum width=1ex,inner sep=.25ex,circle,fill=white,draw] () at (0.50,0.16) {d};
        \node[minimum width=1ex,inner sep=.25ex,circle,fill=white,draw] () at (0.67,0.43) {e};
        \node[minimum width=1ex,inner sep=.25ex,circle,fill=white,draw] () at (0.65,0.08) {f};
        \node[minimum width=1ex,inner sep=.25ex,circle,fill=white,draw] () at (0.53,0.41) {g};
        \node[minimum width=1ex,inner sep=.25ex,circle,fill=white,draw] () at (0.72,0.22) {g};
        \node[minimum width=1ex,inner sep=.25ex,circle,fill=white,draw] () at (0.50,0.515) {h};
        \node[minimum width=1ex,inner sep=.25ex,circle,fill=white,draw] () at (0.74,0.36) {h};
        \node[minimum width=1ex,inner sep=.25ex,circle,fill=white,draw] () at (0.55,0.28) {i};
        \node[minimum width=1ex,inner sep=.25ex,circle,fill=white,draw] () at (0.42,0.67) {i};
    \end{scope}
    \end{tikzpicture}
    \caption{
        3-dimensional model of \gls{vista} with both \gls{bira} positioners installed.
        The static gantry elevation positioner~\encircle{a} moves its probe~\encircle{b} in the plane of the measurement arch~\encircle{c}.
        The static gantry does not move in azimuth, but instead the target is rotated using the turntable~\encircle{d}.
        The moving gantry~\encircle{e} is installed on a semi-circular rail~\encircle{f}, enabling azimuth and elevation positioning fully independently of the static gantry system.
        Pedestals~\encircle{g} raise the center of rotation~\encircle{h} of both gantry arms to 2.27\,m above the floor.
        The flattened quarter circle booms~\encircle{i} enable spherical positioning of two probes on significantly more than a hemisphere.
    }
    \label{fig:mvg_bira_raytraced}
\end{figure}

\Gls{bira} comprises two modular gantry positioners (see \autoref{fig:mvg_bira_raytraced}) enabling almost arbitrary positioning of two probes on the same sphere with a radius of approximately 3\,m and its origin in the focal point 2.27\,m above the turntable in the plane of the multi-probe arch.
Both gantries can be installed independently, enabling single gantry operation also.
The gantry mounting plates are located below the metallic false floor.
When not installed, the respective cutouts in the false floor are covered by metal plates with minimal clearances.
Each gantry is composed of multiple modules that can be installed incrementally, with the first module fastened on the mounting plate.
Precision positioning pins between the modules ensure repeat accuracy between assembly cycles.

\begin{figure}
    \centering
    \begin{tikzpicture}
    \node[anchor=south west,inner sep=0] (image) at (0,0) {\includegraphics[width=1\linewidth]{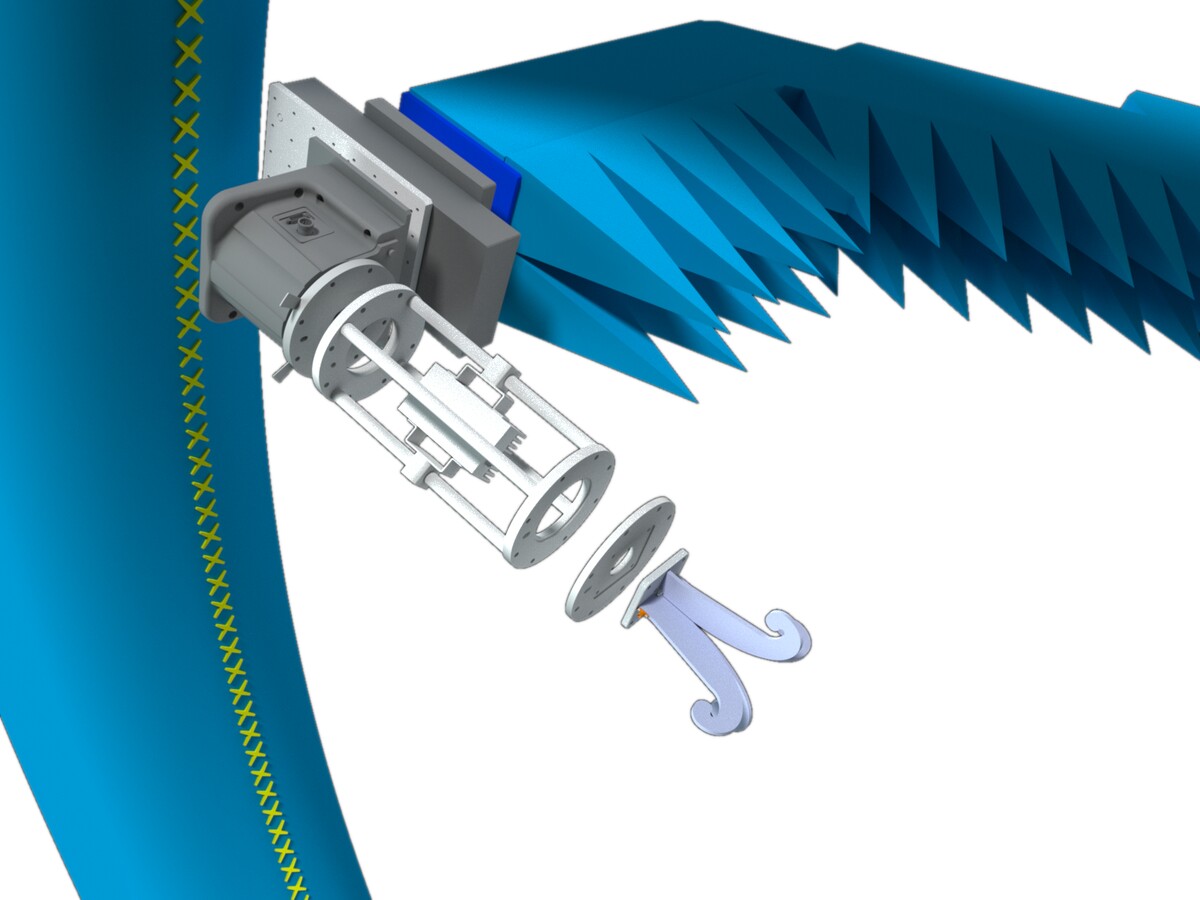}};
    \begin{scope}[x={(image.south east)},y={(image.north west)}]
        \node[minimum width=1ex,inner sep=.25ex,circle,fill=white,draw] () at (0.40,0.86) {a};
        \node[minimum width=1ex,inner sep=.25ex,circle,fill=white,draw] () at (0.10,0.55) {b};
        \node[minimum width=1ex,inner sep=.25ex,circle,fill=white,draw] () at (0.40,0.70) {c};
        \node[minimum width=1ex,inner sep=.25ex,circle,fill=white,draw] () at (0.22,0.70) {d};
        \node[minimum width=1ex,inner sep=.25ex,circle,fill=white,draw] () at (0.37,0.55) {e};
        \node[minimum width=1ex,inner sep=.25ex,circle,fill=white,draw] () at (0.45,0.28) {f};
        \node[minimum width=1ex,inner sep=.25ex,circle,fill=white,draw] () at (0.53,0.23) {g};
        \node[minimum width=1ex,inner sep=.25ex,circle,fill=white,draw] () at (0.60,0.80) {h};
    \end{scope}
    \end{tikzpicture}
    \caption{
        Partially exploded assembly drawing of the static elevation positioner.
        Tip of the flattened quarter circle boom~\encircle{a} in front of the left side of the multi-probe arch~\encircle{b}.
        Radial positioner~\encircle{c} and polarization positioner~\encircle{d} with universal probe flange.
        Exemplary probe~\encircle{e} (microwave amplifier and polarization switch) with same flange as~\encircle{d} on right-hand side.
        Exchangeable adapter disc~\encircle{f} between universal probe flange and antenna~\encircle{g}.
        Microwave absorbers~\encircle{h} are magnetically mounted to facilitate exchanging absorbers according to frequency range.
    }
    \label{fig:mechanic_radial_roll_positioner}
\end{figure}

Relying on the static gantry (see \autoref{fig:mvg_bira_raytraced}\,\encircle{a}), the moving gantry and the turntable, \gls{bira} has a total of eight degrees of freedom, separated into the following axes:
\begin{enumerate}[wide]
    \item \textit{Azimuth:}
        The moving gantry is mounted on a semi-circular rail around the focal point, enabling independent azimuth positioning (see \autoref{fig:mvg_bira_raytraced}\,\encircle{f}).
        The mounting plate of the static gantry is fixed directly to the building foundation.
        With respect to the target on the turntable, the azimuth positioning is realized by rotating the turntable.
    \item \textit{Elevation:}
        Both gantries implement spherical positioning on more than a hemisphere (cf. \autoref{table:sota}) by means of rotating a raised boom in the shape of a flattened quarter circle (see \autoref{fig:mvg_bira_raytraced}\,\encircle{i}).
        The center of rotation of the boom is located in the horizontal plane of the focal point.
        Consequently, the probe at the boom tip always points radially towards the focal point.
        Note that due to this geometry the effective azimuth position of the probe is offset by $\pm$90\textdegree{} relative to the center of rotation.
    \item \textit{Rotation/Polarization:}
        The gantry geometry inherently guarantees that the probe is not rotated around the radial axis towards the focal point when moving in either azimuth or elevation.
        A noteworthy exception is the 180\textdegree{} inversion of the rotation angle when the elevation passes the zenith.
        For the purpose of intentionally rotating the probe, it is fixed to an electronic roll positioner at the tip of the boom (see \autoref{fig:mechanic_radial_roll_positioner}).
        This enables, e.g., dual-polarized measurements with linearly polarized antennas (see \autoref{sec:rf_waveguide}).
    \item \textit{Radius:}
        The roll positioner itself is mounted on an electronic linear positioner aligned with the radial axis (see \autoref{fig:mechanic_radial_roll_positioner}).
        This radial positioner is used, e.g., to ensure probe concentricity despite elastic deformation due to mechanical stress, to account for different probe lengths, or to adjust for frequency-variant antenna phase centers.
        A potential future application could be the augmentation of near-field to far-field transformation via the radial domain~\cite{bevilacqua_eucap23_radial_nearfield_farfield}.
\end{enumerate}

A dedicated laser tracking system was used to repeatedly measure the motion of all axes and to subsequently calibrate and adjust the system.
These measurements have demonstrated a repeatable positioning accuracy for the azimuth axis of 0.02\textdegree{} (0.05\textdegree{} after reassembly) and for the elevation axes of 0.006\textdegree{} (0.02\textdegree{} after reassembly).
The position-dependent elastic deformation from the weight of the positioners and the resulting position error are compensated automatically by the motor controller firmware based on the calibration values.

The firmware is also responsible for mechanical safety:
As both positioners are located predominantly on the surface of the same sphere, collisions are physically possible.
The firmware dynamically predicts an imminent collision and prevents it by automatically enforcing an emergency stop, which is only the last out of several safety measures.
See \autoref{sec:software} for details on the other safety procedures.

The rotating gantry geometry of \gls{bira} enables microwave absorbers on all surfaces that face the target.
This is an advantage over other types of positioners, e.g., robot arms~\cite{jansen23_amta_robot_mmwave_antenna} or traveling trolleys (cf. \autoref{table:sota}), which require an unobstructed rail for their motion.
The absorbers are mounted magnetically to facilitate assembly and disassembly.
Additionally, this enables the use of interchangeable sets of absorbers, each optimized for the respective frequency ranges of interest.

\section{Universal Probe and Transceiver Support}
\label{sec:rf}

Versatility and modularity were primary design requirements for \gls{bira}.
These ensure future extensibility and frequency coverage up to the sub-THz range.
Therefore, no measurement equipment is permanently integrated into the positioners.
Probe flanges on the tips of the rotation positioners enable installation of arbitrary measurement probes, see \autoref{fig:mechanic_radial_roll_positioner}.
The backside of each gantry boom~(\autoref{fig:mvg_bira_raytraced}\,\encircle{i}) supports mounting additional payloads, also within generous size and mass constraints (see \autoref{table:sota}).
For each probe, \gls{bira} provides up to 300\,W of electrical power and \gls{gpio} signals for electronic switching of amplifier gain and polarization.
Additionally, each positioner is equipped with drag chains and empty conduits to enable the mechanically safe and straightforward installation of temporary wiring specific to the probe in use.

\subsection{Integrated VNA Transceivers}
\label{sec:rf_vna}

Microwave measurements based on vector network analysis usually locate \gls{tx} and \gls{rx} ports within one integrated transceiver.
However, bistatic radar reflectivity characterization requires to distribute signals to and from widely separated \gls{tx} and \gls{rx} probes.
\Gls{bira} supports \glspl{vna} with one integrated coaxial cable from each probe to directly adjacent connectors in the anechoic chamber.
A central rotary joint in either elevation positioner allows for a single cable per gantry.

The \gls{tx} cable length sums up to 30\,m, because it passes through the azimuth rail drag chain of the moving positioner.
The \gls{rx} cable spans only 17\,m, as the azimuth position of the static gantry is fixed.
These cable lengths require both active and passive conditioning to enable frequency coverage from 0.5 to 26\,GHz:
Multiple wideband amplifiers compensate for insertion loss, e.g., 60\,dB at 26\,GHz in the \gls{tx} path.
Additionally, passive equalizers flatten the frequency response of the cable to prevent non-linear distortion at lower frequencies, where electronic amplification would otherwise vastly exceed the cable insertion loss.

We employ a Keysight N5222B \gls{vna} in combination with the coaxial cabling described above, to measure the radar reflectivity of static targets or \glspl{aut}.
To enable time gating for the suppression of unwanted reflections, we require a spatial ambiguity limit that exceeds the inner dimensions of our anechoic chamber.
We typically use 10\,MHz frequency steps for a 30\,m range ambiguity.

\subsection{Non-converting RF Probes for VNAs}
\label{sec:rf_nonconverting}

\begin{figure}
    \centering
    \begin{tikzpicture}
    \node[anchor=south west,inner sep=0] (image) at (0,0) {\includegraphics[width=\linewidth]{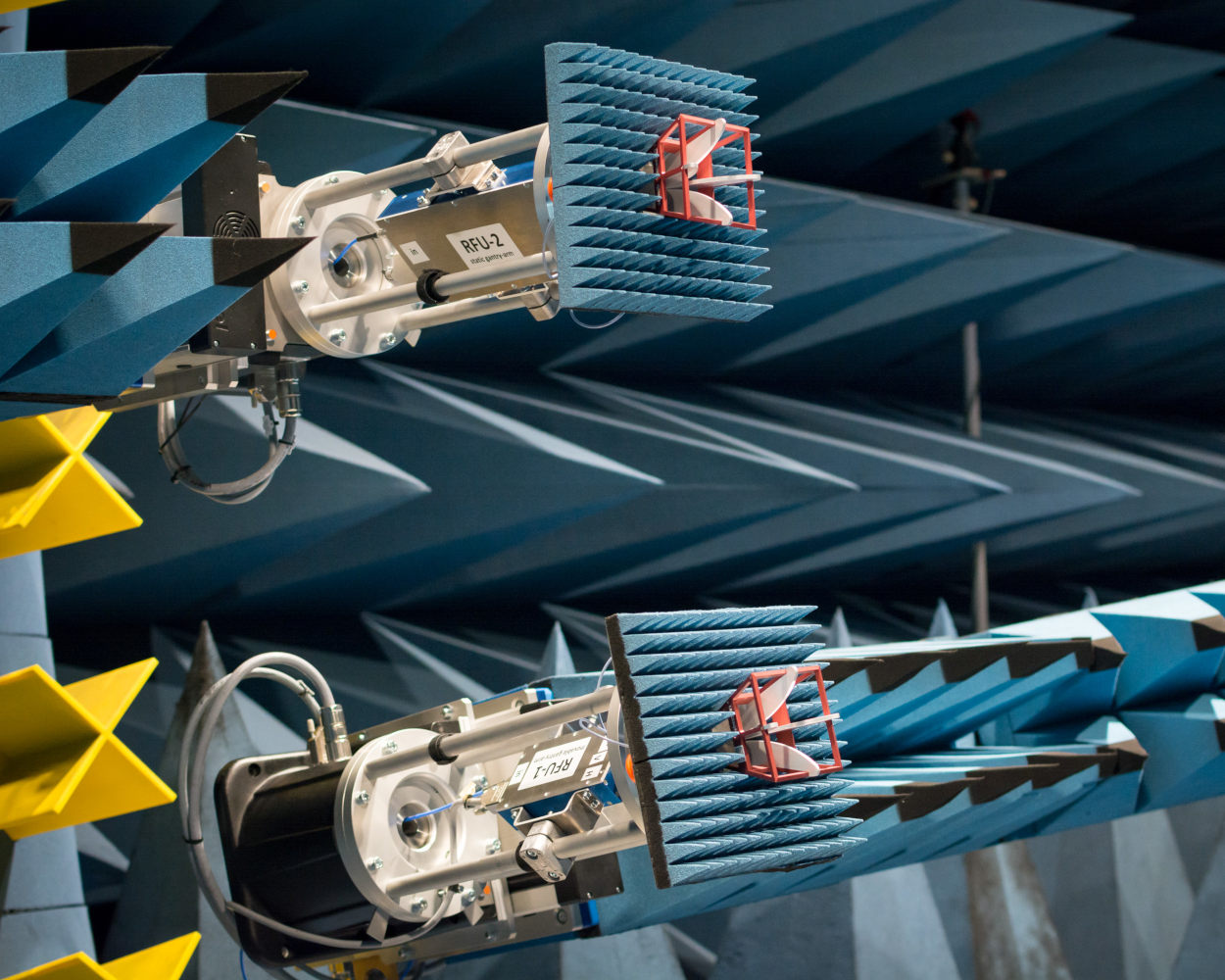}};
    \begin{scope}[x={(image.south east)},y={(image.north west)}]
        \node[minimum width=1ex,inner sep=.25ex,circle,fill=white,draw] () at (0.45,0.23) {a};
        \node[minimum width=1ex,inner sep=.25ex,circle,fill=white,draw] () at (0.40,0.75) {b};
        \node[minimum width=1ex,inner sep=.25ex,circle,fill=white,draw] () at (0.52,0.82) {c};
        \node[minimum width=1ex,inner sep=.25ex,circle,fill=white,draw] () at (0.58,0.25) {c};
    \end{scope}
    \end{tikzpicture}
    \caption{
        Non-converting microwave probes at tip of both gantry arm positioners:
        Tx~\encircle{a} and Rx~\encircle{b} amplifiers housed together with microwave switches for polarization selection.
        RFspin QRH20E dual-polarized antennas~\encircle{c} surrounded by absorbers.
        This configuration is used for \gls{vna} measurements of static targets.
        See \autoref{table:rf} for technical details.
    }
    \label{fig:native_txrx}
\end{figure}

When measuring with integrated transceivers like \glspl{vna}, dual-polarized measurements require electronic switching in combination with the single cable connection between probe and \gls{vna} port.
For our \gls{vna} applications up to 20\,GHz, we rely on a pair of \gls{tx}/\gls{rx} probes that do not perform frequency conversion.
Each probe implements amplification and polarization switching in the frequency range from 0.7 to 20\,GHz.
Both probes include a digital step attenuator for gain adjustment.
Suitable quad-ridged dual-polarized antennas include a pair of Schwarzbeck CTIA0710 and a pair of RFspin QRH20E.
See \autoref{fig:native_txrx} for a picture of both microwave probes mounted on \gls{bira}.

\subsection{Software-defined Parallel Wideband Transceivers}
\label{sec:rf_rfsoc}

Conventional \glspl{vna} are not well-suited for dynamic \gls{isac} measurements.
Their extended sweep times, typically in the order of hundreds of milliseconds, necessitate to step positioners and to measure while stopped, resulting in long measurement durations.
With \gls{bira}, the need to sequentially switch between polarizations exacerbates this issue further.
More importantly, \glspl{vna} are not capable of adequately measuring range and Doppler simultaneously at Nyquist-compliant sampling rates.
While \glspl{vna} can sample Doppler in zero-span mode~\cite{chipengo21access_udoppler_vna_0span,belgiovane17eucap_udoppler_vna_0span}, the resulting lack of range resolution is detrimental to \gls{isac} research, as expected 6G signal bandwidths will enable sensing that resolves the target shape in the time/range-domain.
Consequently, \glspl{vna} are unsuited for simultaneously range- and Doppler-resolved sampling of dynamic target objects that exhibit micro-Doppler effects.

In contrast, modern \glspl{sdr} feature very high instantaneous bandwidths in excess of 1\,GHz.
Excitation signal periods in the order of 1\,µs allow for continuous measurement while the positioner is moving.
This capability dramatically shortens the measurement duration.
Additionally, such short excitation sequences support very high repetition rates and thus Nyquist-compliant sampling of high Doppler frequencies without sacrificing range resolution.

For bistatic measurements, \glspl{sdr} offer another advantage:
\Glspl{vna} are integrated systems and require costly, actively conditioned, long-range signal distribution to and from the separated \gls{tx} and \gls{rx} antennas (cf. \autoref{sec:rf_vna}).
On the contrary, separate \glspl{sdr} can be attached to the \gls{rx} and \gls{tx} probes, realizing a distributed, synchronous measurement architecture.
Generating \gls{tx} and digitizing \gls{rx} signals adjacent to the respective probe allows for inexpensive, lossless, digital signal transfer without practical range limits.
\Glspl{sdr} with parallel \gls{rx} and \gls{tx} channels even facilitate simultaneous, fully polarized measurements without the need to sequentially switch between antenna ports, when relying on orthogonal excitation signals.

We employ two \textit{Xilinx RFSoC ZU47DR} direct \gls{rf} sampling transceivers with a custom \gls{sdr} firmware and software architecture~\cite{engelhardt24_access_rfsoc_arch}.
Each transceiver offers eight \gls{adc} channels and eight \gls{dac} channels operating at sample rates up to 5 or 10\,GSa/s, respectively.
Both \glspl{adc} and \glspl{dac} support sampling in the 2nd and 3rd Nyquist zones with a 6\,GHz upper cutoff frequency.
The RFSoC itself does not require any analog or digital emission suppression methods except for anti-aliasing band-pass filters to attenuate DAC emissions outside of the desired band.
However, non-linear distortions incurred by external front-ends may contribute undesired emissions, e.\,g., due to spectral regrowth.
Outside of shielded environments like BIRA, additional measures may be necessary to suppress potential interference.
A centrally generated, optically distributed, low phase noise clock synchronizes both \glspl{sdr}.
Integrated 100\,Gbit/s Ethernet network interfaces stream samples to and from servers conveniently located outside of the anechoic chamber.
Capable of arbitrary signal generation and sustained recording, the \glspl{sdr} support a wide array of applications beyond bistatic transmissions measurements, e.g., signal and system emulation, hardware-in-the-loop testing, and generic 6G \gls{isac} demonstration~\cite{engelhardt24_access_rfsoc_arch}.

\subsection{Coaxial Quadrature Converters}
\label{sec:rf_coax}

\begin{figure}
    \centering
    \begin{tikzpicture}
    \node[anchor=south west,inner sep=0] (image) at (0,0) {\includegraphics[width=\linewidth]{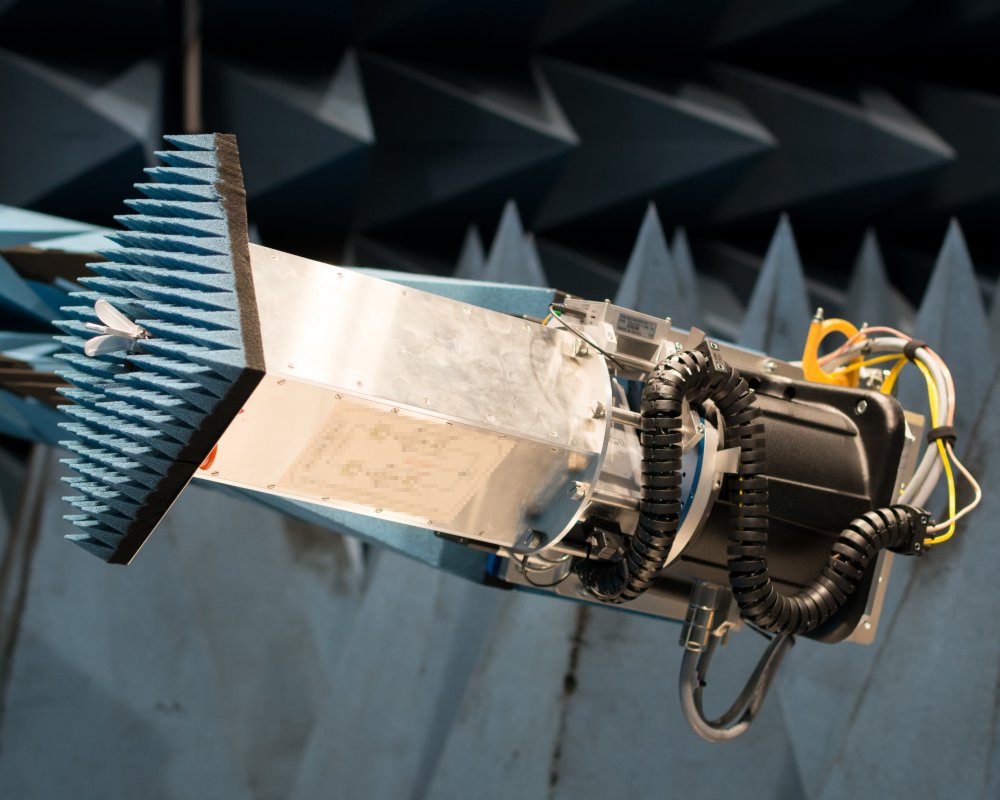}};
    \begin{scope}[x={(image.south east)},y={(image.north west)}]
        \node[minimum width=1ex,inner sep=.25ex,circle,fill=white,draw] () at (0.40,0.57) {a};
        \node[minimum width=1ex,inner sep=.25ex,circle,fill=white,draw] () at (0.12,0.52) {b};
    \end{scope}
    \end{tikzpicture}
    \caption{
        Frequency converter with parallel microwave branches for both polarizations (2 IQ pairs, entirely implemented in coaxial technology).
        These converters are designed primarily for Doppler-resolved measurements with wideband transceivers.
        \Gls{vna} compatibility is provided by IQ splitting/combining via 90° hybrid couplers and microwave switches for polarization selection.
        The picture shows the 20 to 50\,GHz variant~\encircle{a} with RFspin QRH50E antenna~\encircle{b}.
        A similar converter pair is available for 5 to 20\,GHz with RFspin QRH20E antenna.
        See \autoref{table:rf} for technical details.
    }
    \label{fig:rf_converter_coax}
\end{figure}

\begin{table*}[ht!]
\centering
\caption{Typical performance parameters of BIRA microwave probes.}
\label{table:rf}
\scriptsize
\renewcommand{\arraystretch}{1.1}
\begin{tabular}{ |l|r|r|r|r|r|r|  }
\hline
    &	non-converting probe	&	IQ converter	&	IQ converter	&	WR-12 converter	&	WR-6.5 converter	&	WR-4.3 converter \\
\hline
\hline
frequency range	&	0.7\,...\,20\,GHz	&	5\,...\,20\,GHz	& 20\,...\,50\,GHz	&	60\,...\,90\,GHz	&	110\,...\,170\,GHz	&	170\,...\,260\,GHz \\
\hline
RF filters & -- & -- & -- & 76\,...\,81\,GHz & 140\,...\,148\,GHz & 200\,...\,260\,GHz \\
 &    &    &    & 60\,...\,90\,GHz & 110\,...\,140\,GHz &  \\
 &    &    &    & 70\,...\,90\,GHz  & 140\,...\,170\,GHz &  \\
\hline
IF range & N/A & 0\,...\,6\,GHz & 0\,...\,23\,GHz & \,...\,12\,GHz & \,...\,17\,GHz & \,...\,26\,GHz \\
\hline
LO range & N/A & 5\,...\,20\,GHz & 9\,...\,25\,GHz & 15\,...\,22\,GHz & 18.3\,...\,28.3\,GHz & 14.2\,...\,21.6\,GHz \\
\hline
Tx $P_\text{in1dB}$ & --13\,dBm & 10\,dBm & 1\,dBm & --20\,dBm & --20\,dBm & --20\,dBm \\
\hline
Tx $P_\text{out1dB}$ & 22\,dBm & 17\,dBm & 13\,dBm & 6\,dBm & 2\,dBm & 5\,dBm \\
\hline
Tx gain & 35\,dB & 7\,dB & 12\,dB & 26\,dB & 22\,dB & 25\,dB \\
\hline
Rx noise figure & 4\,dB & 6\,dB & 7\,dB & 6\,dB & 7\,dB & 8\,dB \\
\hline
Rx gain & 30\,dB & 26\,dB & 30\,dB & 38\,dB & 34\,dB & 53\,dB \\
\hline
Rx $P_\text{in1dB}$ & --15\,dBm & --14\,dBm & --16\,dBm & --13\,dBm & --18\,dBm & --28\,dBm \\
\hline
polarization &  dual, linear & dual, linear & dual, linear & single, linear & single, linear & single, linear \\
selection  &  RF switch & parallel (2$\times$IQ pair)  & parallel (2$\times$IQ pair) & mechanical rotation & mechanical rotation & mechanical rotation \\
\hline
antennas & Schwarzbeck CTIA 0710 & RFspin QRH20E & RFspin QRH50E & 25\,dBi, 9$^\circ$\hspace*{0.5em} & 25\,dBi, 9$^\circ$\hspace*{0.5em} & 25\,dBi, 10$^\circ$ \\
    & RFspin QRH20E &  &  & 20\,dBi, 14$^\circ$ & 20\,dBi, 17$^\circ$ &  \\
    &  &  &  & 15\,dBi, 30$^\circ$ & 15\,dBi, 33$^\circ$ &  \\
    &  &  &  & 10\,dBi, 55$^\circ$ &  &  \\
\hline
\end{tabular}
\end{table*}

The upper cutoff frequency of the \gls{sdr} transceivers necessitates frequency converters for measurements above 6\,GHz.
Up to approximately 67\,GHz, coaxial technology offers straightforward and affordable dual-polarized antennas and converters.
In particular, quadrature mixers available for this frequency range provide flexible frequency coverage through inherent image suppression without external filters and \gls{if} bandwidths of several GHz (cf. \autoref{table:rf}).
Their quadrature \gls{if} interface doubles the native instantaneous bandwidth of multi-port transceivers, achieving 4\,GHz bandwidth in combination with our \glspl{sdr}.
We designed and assembled two pairs of quadrature frequency converters in coaxial technology:
One up-converter (\gls{tx}) and one down-converter (\gls{rx}) each for the frequency ranges from 5 to 20\,GHz and from 18 to 50\,GHz.
All converters use dual linearly polarized antennas and fully parallel microwave branches for both polarizations, requiring four \gls{adc} or \gls{dac} ports per converter, respectively.
Technical details are summarized in \autoref{table:rf}.
The components are integrated into a shielded case with microwave absorbers magnetically mounted on the front and a flange compatible with \gls{bira}, see \autoref{fig:rf_converter_coax}.
Phase-coherent operation of \gls{tx} and \gls{rx} is ensured by \gls{lo} sharing:
Within the \gls{bira} setup, the \gls{lo} is fiber-optically distributed from a common signal generator (Keysight EXG N5173B).

The quadrature probes support \glspl{vna} through analog signal adapters:
The IQ channels are externally combined via 90° hybrid couplers and polarization selection is carried out via microwave switches that are controlled remotely with \gls{gpio} signals.
The integrated cabling (cf. \autoref{sec:rf_vna}) then routes the resulting single \gls{if} signal between each probe and the \gls{vna}.

\subsection{Waveguide Converters}
\label{sec:rf_waveguide}

\begin{figure}
    \centering
    \begin{tikzpicture}
    \node[anchor=south west,inner sep=0] (image) at (0,0) {\includegraphics[width=\linewidth]{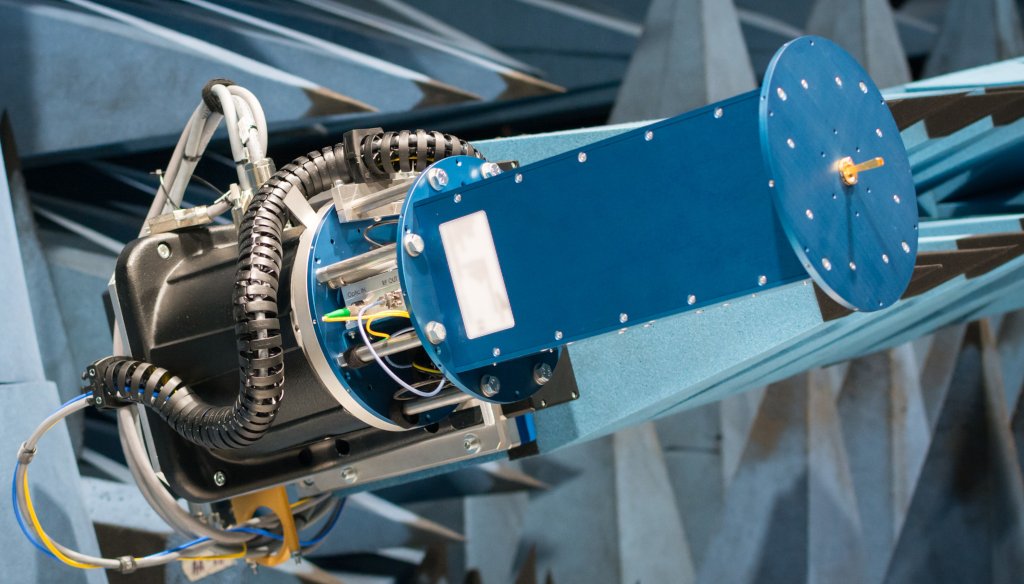}};
    \begin{scope}[x={(image.south east)},y={(image.north west)}]
        \node[minimum width=1ex,inner sep=.25ex,circle,fill=white,draw] () at (0.60,0.60) {a};
        \node[minimum width=1ex,inner sep=.25ex,circle,fill=white,draw] () at (0.835,0.64) {b};
        \node[minimum width=1ex,inner sep=.25ex,circle,fill=white,draw] () at (0.35,0.55) {c};
        \node[minimum width=1ex,inner sep=.25ex,circle,fill=white,draw] () at (0.26,0.35) {d};
    \end{scope}
    \end{tikzpicture}
    \caption{
        Frequency converter~\encircle{a}, implemented in waveguide technology.
        Here, the frontal absorbers have been removed to make the minuscule horn antenna~\encircle{b} visible.
        \Gls{lo} signals are distributed fiber-optically.
        The opto-electrical converter~\encircle{c} is stacked between the \gls{bira} flange and the frequency converter case.
        Cables are routed through a flexible drag chain~\encircle{d} to safely enable polarization selection via mechanical rotation.
        The photograph depicts the WR-6.5 waveguide band variant (110 to 170\,GHz).
        Similar converters are available as well for WR-12 (60 to 90\,GHz) and WR-4.3 (170 to 260\,GHz).
        See \autoref{table:rf} for technical details.
    }
    \label{fig:rf_converter_waveguide}
\end{figure}

Targeting frequencies above 67\,GHz requires moving from coaxial to waveguide components.
Three pairs of converters are available for the waveguide bands WR-12 (60 to 90\,GHz), WR-6.5 (110 to 170\,GHz), and WR-4.3 (170 to 260\,GHz), see \autoref{fig:rf_converter_waveguide}.
Within these frequency ranges, individual bands can be selected by exchanging band-pass filters.
The trade-off between antenna directivity and angular field-of-view is addressed by multiple antenna pairs.
All parameters are documented in \autoref{table:rf}.
All waveguide converters are implemented single-polarized.
Dual-polarized measurements are realized by remote-controlled mechanical rotation of the probes.
As with the IQ converters, phase-coherent operation is ensured by fiber-optic \gls{lo} distribution and by applying \gls{lo} multiplication.
Hardware integration and supply of the waveguide converters was provided by the contractor \textit{bsw TestSystems \& Consulting}.

\section{Software}
\label{sec:software}

The operation of the eight mechanical \gls{bira} axes is not straightforward. 
Firstly, its machine coordinates differ from the bistatic angles $\azi$, $\eli$, $\azo$, and $\elo$ of both probes with respect to the target.
The mapping from bistatic angles to machine coordinates is not unique.
Secondly, collisions, although prevented by firmware, are hypothetically possible, due to the fact that both gantry positioners move on the same sphere around the focal point.
This necessitates careful planning and verification of measurement trajectories, i.e., consecutive lists of positioner waypoints, to ensure safe and reliable operation.
The trajectories must pre-emptively circumnavigate gantry collisions while also minimizing the extent of detours incurred by this process.

The \gls{bira} software suite comprises two primary components:
A mechanically exact digital twin and a hardware abstraction layer combined with a generic \gls{api}.

\subsection{Interactive Digital Twin}
\label{sec:software_digtwin}

\begin{figure}
    \centering
    \begin{tikzpicture}
    \node[anchor=south west,inner sep=0] (image) at (0,0) {\includegraphics[width=1\linewidth]{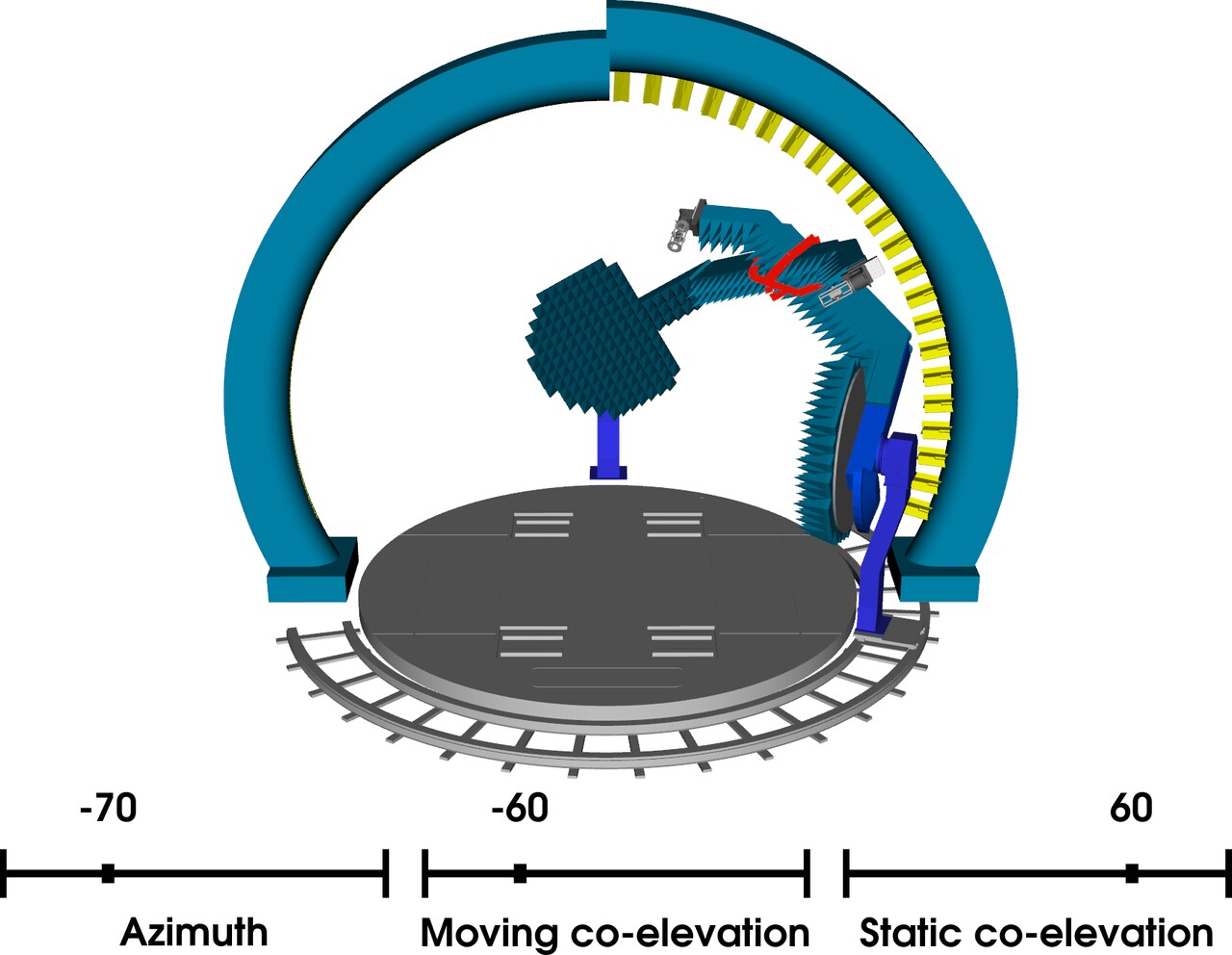}};
    \begin{scope}[x={(image.south east)},y={(image.north west)}]
        \node[minimum width=1ex,inner sep=.25ex,circle,fill=white,draw] () at (0.62,0.65) {a};
    \end{scope}
    \end{tikzpicture}
    \caption{
        Screenshot of the \gls{bira} digital twin depicting a mechanically impossible state  with both gantries colliding.
        This instance of the digital twin with enabled GUI allows interactive control of the gantry positions through slider widgets at the bottom of the screen.
        Bounding boxes with 10\,cm safety clearance are used for collision detection but not rendered.
        Only bounding box polygon intersections, i.e., likely collisions, are drawn in red~\encircle{a}.
    }
    \label{fig:digital_twin}
\end{figure}

We implemented the digital twin using the Python programming language and the Visualization Toolkit (VTK) computer graphics library.
The twin relies on the \gls{cad} models of \gls{bira} to implement its geometrically exact replica.
Additionally, a simplified bounding box is used for collision detection with a safety clearance of 10\,cm.

The digital twin includes an optional, interactive graphical user interface (GUI) (see \autoref{fig:digital_twin}).
The GUI visualizes the machine coordinate system, which facilitates system familiarization for first-time users and accelerates measurement trajectory planning significantly.
We also employ the GUI for safe interactive control of \gls{bira} by visualization of movements prior to execution.

\begin{figure}
    \centering
    \begin{tikzpicture}
\pgfdeclarelayer{foreground}
\pgfsetlayers{main,foreground}
\begin{axis}[
        scale mode=scale uniformly,
        scale uniformly strategy=units only,
        width=\linewidth,
        enlargelimits=false,
        axis on top,
        tick align=outside,
        tick pos=left,
        xlabel={Static gantry co-elevation (deg)},
        ylabel={Moving gantry co-elevation (deg)},
        xmajorgrids,
        ymajorgrids,
        xtick distance=15,
        ytick distance=15,
        x tick label style={rotate=90,anchor=east}
    ]
    \addplot graphics[xmin=-115, xmax=115, ymin=-115, ymax=115] {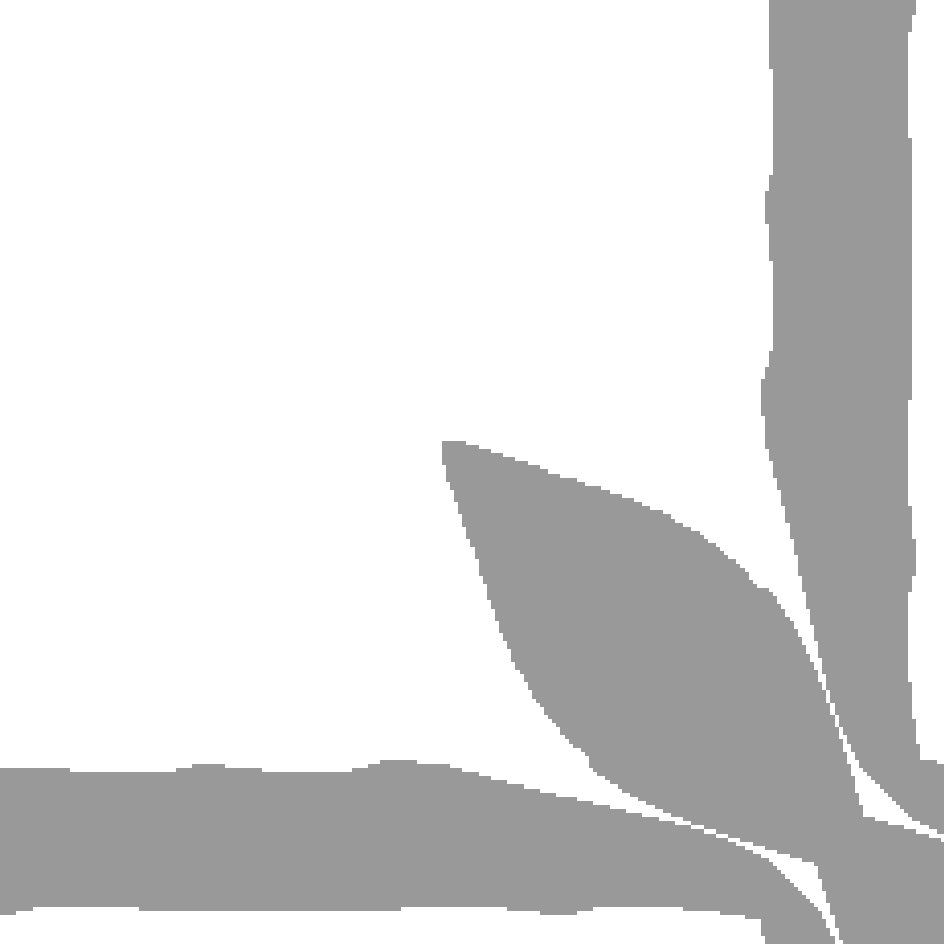};
    \begin{pgfonlayer}{foreground}
        \node[minimum width=1ex,inner sep=.25ex,circle,fill=white,draw] () at (60,-60) {a};
    \end{pgfonlayer}
\end{axis}
\end{tikzpicture}
    \caption{
        2-dimensional slice of collision table generated by digital twin for the machine coordinate azimuth angle \mbox{--70\textdegree{}}.
        Shaded areas denote angle combinations where bounding boxes collide.
        The gantry positions of the collision rendered in \autoref{fig:digital_twin} are marked by \encircle{a}.
    }
    \label{fig:collision_table}
\end{figure}

The digital twin is essential for mechanical safety.
Being realized as a multi-purpose library with a flexible \gls{api}, we also use it to compute a collision table containing all possible permutations of the three main mechanical axes of \gls{bira}:
Moving gantry azimuth and co-elevation as well as static gantry co-elevation.
The range of motion of the remaining axes precludes them from any contribution to possible collisions.
The table contains approximately 10~million entries at 1\textdegree{} angular resolution, see \autoref{fig:collision_table} for an exemplary section of the table.
The simplified bounding box approach used to compute intersections for the collision table only requires a half-hour computation time, facilitating generation and use of alternative collision tables for non-standard probes or large targets.
A default table for cylindrical probes as depicted in \autoref{sec:rf} is used by the firmware and -- possibly alternative -- tables are used by the software to implement failsafe, multi-layered collision prevention.

\subsection{Hardware Abstraction Layer and Application Programming Interface~(API)}
\label{sec:software_api}

The primary design goal of \gls{bira} was universal usability, requiring a generic \gls{api} not limited to specific measurement types.
This user-facing \gls{api} must satisfy the following requirements in the order of priority:
Mechanical safety, efficiency, and user-friendliness.
All requirements rule out granting users direct access to the positioner hardware.
Instead, we opted for a safety enforcing but straightforward abstraction layer:
Users submit their measurement trajectory as a plain text file that contains a list of consecutive axis positions.
Qualified staff members select additional axis motion parameters, e.g., velocity, acceleration, and deceleration.
For the verification of user-supplied trajectories, we employ a kinematic model of all eight axes, which computes their velocity-, acceleration-, and jerk-limited motion.
The model-predicted motion is checked against the configured collision table computed by the digital twin.
We also use the kinematic model to calculate exact measurement duration and to assess trajectory efficiency, e.g., to identify unnecessary detours.
After successful verification, trajectories are cleared for long-term, automated, unattended measurements over night or weekends.

Four bistatic angles and thus four primary degrees of freedom result in typical measurement times in the range from several hours up to multiple days.
This implies that even minor efficiency gains result in significant time savings.
Our measurement software interfaces the turntable, the gantry positioners, and the \gls{rf} measurement devices, e.g., a \gls{vna} or \gls{sdr}, via an Ethernet local area network.
The involved industrial motor controllers have network response latencies around 100\,ms.
Therefore, we implemented fully parallelized network remote control using the Python \textit{asyncio} library and asynchronous coroutines.
We batch movement commands for multiple axes into single network requests and optimize for successful command execution by deferring error checks until after all movement commands were issued.
This way, we achieve a total remote control overhead of only 100\,ms per step.
In contrast, multiple sequential network requests would add up query latencies, accumulating up to to several hours for long-term measurements.

Like \gls{bira} itself, our software also supports arbitrary, user-defined measurement applications.
Whereas this requires the integration of custom user code into the measurement software program flow, user-provided program code and data must remain strictly isolated from the positioner hardware for safety reasons.
We realized this through a restricted callback \gls{api}.
For stepped measurements, our software calls a user-provided function after all axes have come to standstill at their respective target position.
For continuously moving measurements, our software runs an asynchronous user-provided coroutine in parallel to continuously orchestrate all motions.
The user code has read-only access to all states, e.g., current position and velocity, but cannot issue any commands.
To facilitate the development of user code, the measurement \gls{api} supports offline use without hardware access, partially simulating positioner behavior.
Standard implementations for \gls{vna} and \gls{sdr} measurements are available and can be used without modification for most measurement applications.

\section{Measurement Examples}
\label{sec:examples}

Subsequently, we exemplify the functionality and versatility of \gls{bira} with multiple measurements.
First, \autoref{sec:examples_rcs_sphere} validates the correctness of bistatic reflectivity measurements by using a metal sphere as a canonical reference object and by comparing its measured \gls{rcs} with analytical solutions as well as simulation results.
\autoref{sec:sphere_time_domain} analyzes the metal sphere reflectivity in time-domain in order to illustrate the challenges incurred by wideband reflectivity measurement, e.g., the necessity of antenna calibration and deconvolution for accurate sensing.
Please note that the calibration methods presented here are for illustrative purposes only.
A thorough description of bistatic measurement calibration is beyond the scope of this paper.
Subsequently, \autoref{sec:examples_rcs_double_sphere} investigates an extended target that can be easily modeled by expanding the previous example with a second metal sphere.
Then, \autoref{sec:examples_micro_doppler} showcases a real-world example:
The wideband micro-Doppler reflectivity of a moving pedestrian \gls{ncap} dummy in combination with changing bistatic geometry.
Finally, \autoref{sec:examples_antenna} exemplarily validates an antenna measurement with \gls{bira} by comparison with two proven antenna measurement systems.

\subsection{System Verification: RCS of Metal Sphere}
\label{sec:examples_rcs_sphere}\begin{figure}
    \centering
    \begin{tikzpicture}
        \node[anchor=south west,inner sep=0] (image) at (0,0) {\includegraphics[width=1\linewidth]{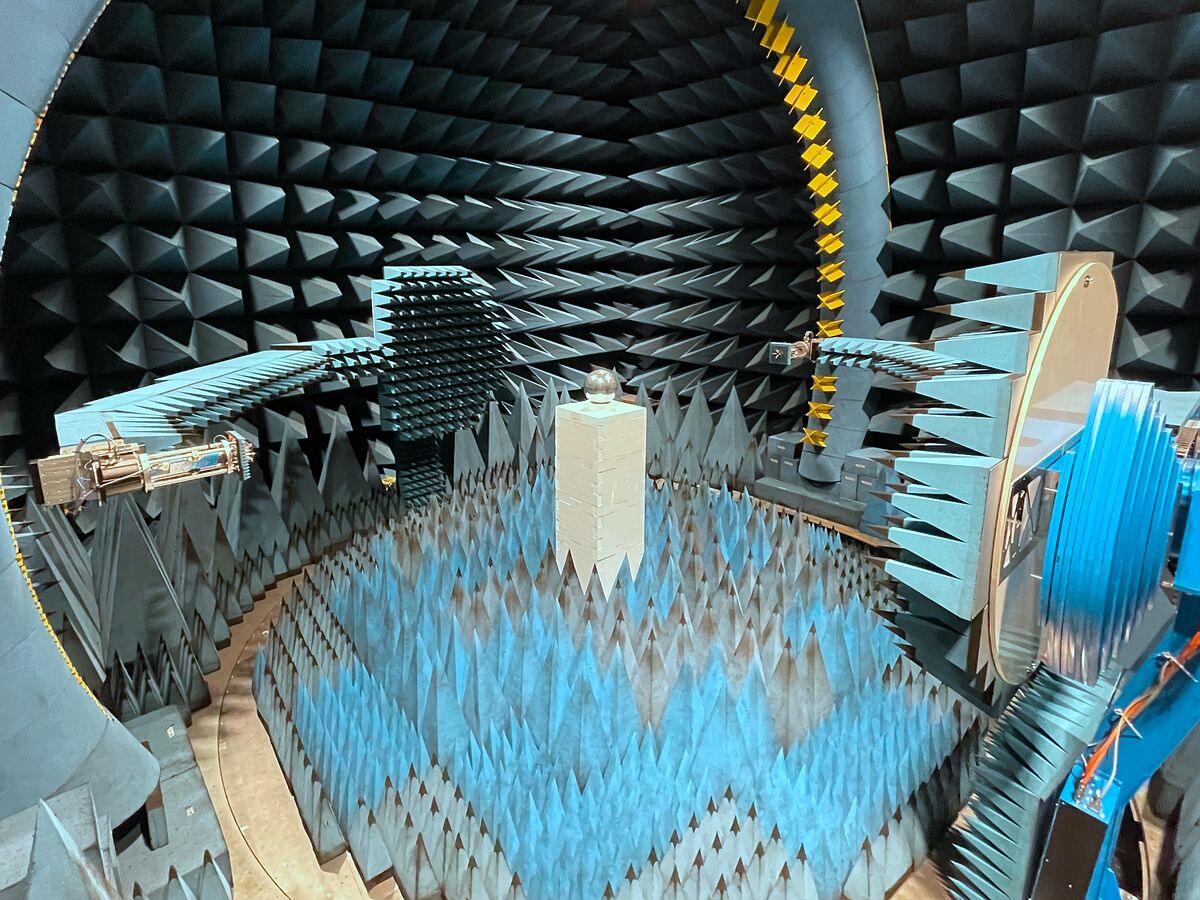}};
        \begin{scope}[x={(image.south east)},y={(image.north west)}]
            \node[minimum width=1ex,inner sep=.25ex,circle,fill=white,draw] () at (0.50,0.50) {a};
            \node[minimum width=1ex,inner sep=.25ex,circle,fill=white,draw] () at (0.05,0.46) {b};
            \node[minimum width=1ex,inner sep=.25ex,circle,fill=white,draw] () at (0.65,0.65) {c};
        \end{scope}
    \end{tikzpicture}
    \caption{
        Bistatic \gls{rcs} measurement setup of a metallic sphere with \spherebig\ diameter on a Styrodur pillar~\encircle{a}.
        On the static gantry, the \gls{rx}~\encircle{b} stays at $\elo$~=~90°.
        On the moving gantry, the \gls{tx}~\encircle{c} realizes a horizontal cut.
    }
    \label{fig:rcs_sphere_setup}
\end{figure}

\begin{figure}
\centering
\begin{tikzpicture}[font=\scriptsize,spy using outlines={circle, magnification=4, size=2cm, connect spies}]
\begin{polaraxis}[
   width=0.87\columnwidth,
   title style={align=center},
   title = {$\sigma_{\text{sphere\,$\varnothing$30\,cm}}(\azi,\eli=100^\circ,\azo=0°,\elo=90°,f=6\,\text{GHz})$ in dBsm\\$\varphi$ polarization},
   title = {$\sigma_\text{sphere}(\azo\!-\!\azi)$ in dBsm\\$\az$ polarization},
   title = {$\az$ polarization},
   every axis title/.style={below right,at={(0.27,0.9)},draw=black,fill=white},
   xticklabel=$\pgfmathprintnumber{\tick}^\circ$,
   xticklabels={0$^\circ$,30$^\circ$,60$^\circ$,90$^\circ$,120$^\circ$,150$^\circ$,180$^\circ$,210$^\circ$,240$^\circ$,270$^\circ$,300$^\circ$,330$^\circ$},
   xtick={0,30,...,330},
   xmin=0, xmax=540,
   ytick={-40,-30,...,20},
   yticklabels={--40,,--20,,0,,20},
   ymin=-40, ymax=20,
   x coord trafo/.code=\pgfmathparse{#1+180},
   x coord inv trafo/.code=\pgfmathparse{#1-180},
   y coord trafo/.code=\pgfmathparse{#1+40},
   y coord inv trafo/.code=\pgfmathparse{#1-40},
   legend style={
  at={(current bounding box.south-|current axis.south)},
  yshift=-0.0cm,
  anchor=north,
  legend columns=-1
},
   legend entries={BIRA\hspace*{1em}, HFSS\hspace*{1em},analytic\hspace*{1em}},
]
\addplot+[mark=none,line join=bevel,blue,densely dashdotted] table[x=angle,y=H] {Figures/measurements/1sphere/RCS_sphere_30cm_6ghz_bira.txt};
\addplot+[mark=none,line join=bevel,black,solid] table[x=angle,y=H] {Figures/measurements/1sphere/RCS_sphere_30cm_6ghz_HFSS.txt};
\addplot+[mark=none,line join=bevel,red,densely dashdotdotted] table[x=angle,y=H] {Figures/measurements/1sphere/RCS_sphere_30cm_6ghz_anal_flipped.txt};
\coordinate (spy11) at (axis cs:210,-10);
\coordinate (mag11) at (axis cs:230,15);
\coordinate (spy12) at (axis cs:150,-10);
\coordinate (mag12) at (axis cs:130,15);
\end{polaraxis}
\spy[gray, size=2.0cm, magnification=3] on (spy11) in node[fill=white] at (mag11);
\spy[gray, size=2.0cm, magnification=3] on (spy12) in node[fill=white] at (mag12);
\end{tikzpicture}
\begin{tikzpicture}[font=\scriptsize,spy using outlines={circle, magnification=4, size=2cm, connect spies}]
\begin{polaraxis}[
   width=0.87\columnwidth,
   title = {$\el$ polarization},
   every axis title/.style={below right,at={(0.27,0.9)},draw=black,fill=white},
   xticklabel=$\pgfmathprintnumber{\tick}^\circ$,
   xticklabels={0$^\circ$,30$^\circ$,60$^\circ$,90$^\circ$,120$^\circ$,150$^\circ$,180$^\circ$,210$^\circ$,240$^\circ$,270$^\circ$,300$^\circ$,330$^\circ$},
   xtick={0,30,...,330},
   xmin=0, xmax=540,
   ytick={-40,-30,...,20},
   yticklabels={--40,,--20,,0,,20},
   ymin=-40, ymax=20,
   x coord trafo/.code=\pgfmathparse{#1+180},
   x coord inv trafo/.code=\pgfmathparse{#1-180},
   y coord trafo/.code=\pgfmathparse{#1+40},
   y coord inv trafo/.code=\pgfmathparse{#1-40},
]
\addplot+[mark=none,line join=bevel,blue,densely dashdotted] table[x=angle,y=V] {Figures/measurements/1sphere/RCS_sphere_30cm_6ghz_bira.txt};
\addplot+[mark=none,line join=bevel,black,solid] table[x=angle,y=V] {Figures/measurements/1sphere/RCS_sphere_30cm_6ghz_HFSS.txt};
\addplot+[mark=none,line join=bevel,red,densely dashdotdotted] table[x=angle,y=V] {Figures/measurements/1sphere/RCS_sphere_30cm_6ghz_anal_flipped.txt};
\coordinate (spy21) at (axis cs:197,-5);
\coordinate (mag21) at (axis cs:230,15);
\coordinate (spy22) at (axis cs:163,-5);
\coordinate (mag22) at (axis cs:130,15);
\end{polaraxis}
\spy[gray, size=2.0cm, magnification=3] on (spy21) in node[fill=white] at (mag21);
\spy[gray, size=2.0cm, magnification=3] on (spy22) in node[fill=white] at (mag22);
\end{tikzpicture}
\caption{Bistatic RCS in dBsm (measured with \gls{bira}, simulated with HFSS, and analytically calculated) of a metallic sphere with \spherebig\ diameter at 6\,GHz along bistatic azimuth angle $\azi-\azo$, fixed co-elevation $\el=90$°.}
\label{fig:rcs_sphere}
\end{figure}

\begin{figure*}
    \centering
    \begin{tikzpicture}
\pgfdeclarelayer{foreground}
\pgfsetlayers{main,foreground}
\begin{axis}[
        domain=10:243,
        colorbar,
        colormap/jet,
        colorbar style={
            ylabel={Reflectivity (dBsm)},
            ytick={-80,-60,...,10},
            yticklabel style={
                text width=1.5em,
                align=right
            }
        },
        width=0.89\linewidth,
        height=4cm,
        enlargelimits=false,
        axis on top,
        tick align=outside,
        tick pos=left,
        xtick pos=top,
        title={Normalized},
        every axis title/.style={below left,at={(1,1)},draw=black,fill=white},
        ylabel={Normalized range (m)},
        xtick distance=15,
        ytick distance=0.1,
        x tick label style={rotate=90,anchor=west}
    ]
    \addplot[point meta min=-80, point meta max=20] graphics[xmin=10, xmax=243, ymin=-0.5, ymax=0.5] {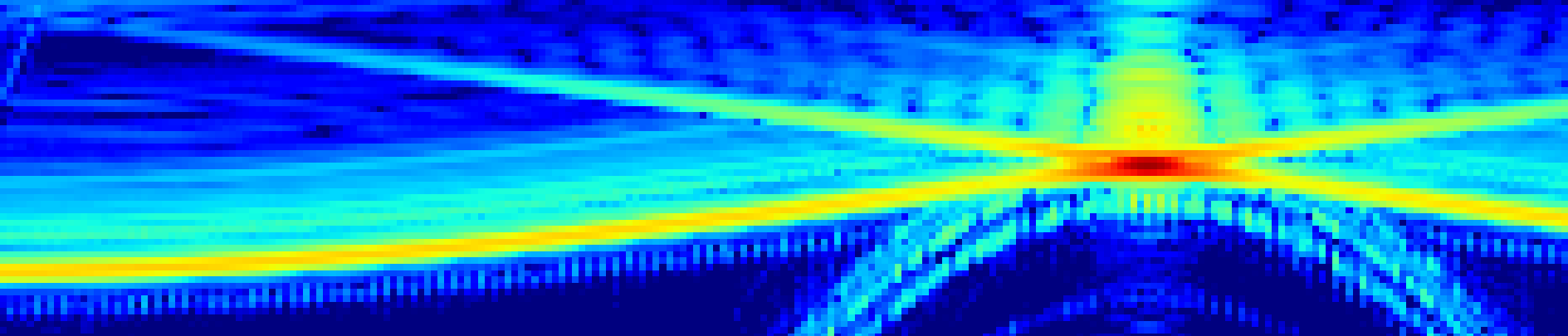};
    \addplot[mark=o, mark repeat = 6, mark phase = 3,samples=234,only marks] {abs((x-180)/360*pi*0.3048) + 0.008};
    \addplot[mark=diamond, mark repeat = 6, mark phase = 6,samples=234,only marks] {2*((2.9 - 0.3048/2*abs(cos(x/2)))^2 + (0.3048/2*abs(sin(x/2)))^2)^0.5 - 2*2.9};
\end{axis}
\end{tikzpicture}

\begin{tikzpicture}
\pgfdeclarelayer{foreground}
\pgfsetlayers{main,foreground}
\begin{axis}[
        domain=10:243,
        colorbar,
        colormap/jet,
        colorbar style={
            ylabel={Reflectivity (dBsm)},
            ytick={-80,-60,...,10},
            yticklabel style={
                text width=1.5em,
                align=right
            }
        },
        width=0.89\linewidth,
        height=4cm,
        enlargelimits=false,
        axis on top,
        tick align=outside,
        tick pos=left,
        xtick pos=bottom,
        title={Calibrated},
        every axis title/.style={below left,at={(1,1)},draw=black,fill=white},
        xlabel={Bistatic azimuth angle $\azi-\azo$ (deg)},
        ylabel={Normalized range (m)},
        xtick distance=15,
        ytick distance=0.1,
        x tick label style={rotate=90,anchor=east}
    ]
    \addplot[point meta min=-80, point meta max=20] graphics[xmin=10, xmax=243, ymin=-0.5, ymax=0.5] {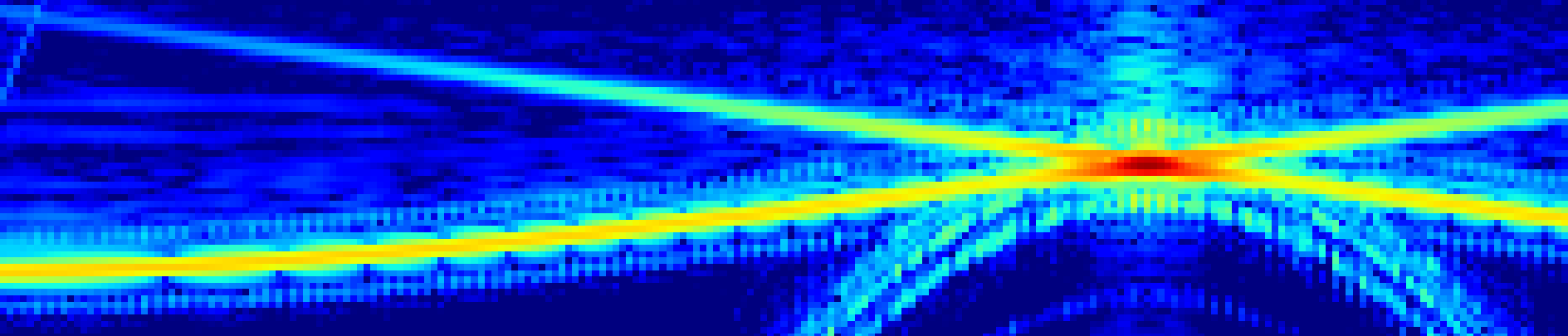};
    \addplot[mark=o, mark repeat = 6, mark phase = 3,samples=234,only marks] {abs((x-180)/360*pi*0.3048) + 0.008};
    \addplot[mark=diamond, mark repeat = 6, mark phase = 6,samples=234,only marks] {2*((2.9 - 0.3048/2*abs(cos(x/2)))^2 + (0.3048/2*abs(sin(x/2)))^2)^0.5 - 2*2.9};
\end{axis}
\end{tikzpicture}
    \caption{
        Range-resolved, bistatic reflectivity of metal a sphere with diameter \spherebig\ measured from 2 to 18\,GHz in the horizontal plane with horizontal polarization.
        Top plot without and bottom plot with antenna calibration and deconvolution using the algorithm derived in~\cite{andrich25_eucap_antenna_deconvolution_radar_refl}.
        Range normalized to center of sphere and focal point.
        The close path (diamonds) represents the tangential reflection and the distant path (circles) the creeping wave diffracted around the sphere as illustrated in \autoref{fig:sphere_bistatic_paths}.
        Both paths coincide with geometric models indicated by circles and diamonds.
        The repetitive pattern visible around the reflected path is caused by sampling of Hann window nulls and sidelobes.
    }
    \label{fig:sphere_impresp}
\end{figure*}

As a baseline verification of \gls{bira} in its mainly intended purpose for bistatic radar reflectivity measurements, we compare measured data of a metallic sphere with computer simulations and analytic calculation.
The \gls{rcs} evaluation was performed according to~\cite{schwind20_amta_rcs_vru_scaled}, incorporating an additional measurement of a reference object with known \gls{rcs}.
In this case, two metal spheres with different diameters were measured, one as a test object (diameter \spherebig) and the other as a reference object (diameter \spheresmall).
Both positioners and the sphere were located in the horizontal plane as pictured in \autoref{fig:rcs_sphere_setup}.
The \gls{rx} remained static and the \gls{tx} was stepped to realize bistatic angles $\azo - \azi$ from --10\textdegree{} to +170\textdegree{}.
RFspin QRH20E antennas and the \gls{vna} setup described in \autoref{sec:rf_vna} were used to measure from 2 to 18\,GHz in 10\,MHz steps.

Postprocessing involved background subtraction (measurement without target) from the target measurement and time-domain gating to reduce parasitic reflections.

To validate the measurement results, we look at both the analytical solution~\cite{balanis_2012_electromagnetics} and electromagnetic simulations with ANSYS HFSS SBR+ as reference.
The analytical \gls{rcs} is valid under far-field conditions, whereas the simulations took into account the actual near-field conditions.
\autoref{fig:rcs_sphere} shows measurement data and both simulations at fixed co-elevation $\el$\,=\,90°.
The calculated and simulated curves show good agreement, with up to 0.7\,dB difference for $\el$ polarization and 1.3\,dB for $\az$ polarization in the backscatter region (0°\ldots $\pm$90°).
In the forward scattering region, small deviations in position and depth of the minima appear, which is due to the near-field (HFSS) vs. far-field (analytic) assumption.
Comparing the simulations with the measurement data, we find an excellent matching in the values and the characteristic patterns.

This convincing agreement of the measured data with the expected values verifies the correct functioning of \gls{bira}.

\begin{figure}
    \centering
    \begin{tikzpicture}[font=\scriptsize]
\pgfdeclarelayer{foreground}
\pgfsetlayers{main,foreground}
\begin{groupplot}[
    group style={
        group size=2 by 1,
        horizontal sep=2mm,
        y descriptions at=edge left,
    },
    ylabel={Normalized power (dB)},
    height=5cm,
    scale only axis,
    ymin=-100,
    ymax=0,
    grid=major
]

\nextgroupplot[
    xmin=-0.1,
    xmax=0.5,
    width=0.175\columnwidth
]
\addplot+[line join=bevel,blue,mark=o,mark repeat=15,mark phase=6] table[x=range_m,y=amplitude_dB] {Figures/measurements/1sphere_time/01_antcal_raw_24_metal.txt};
\addplot+[line join=bevel,red,mark=diamond,mark repeat=15,mark phase=6] table[x=range_m,y=amplitude_dB] {Figures/measurements/1sphere_time/01_antcal_raw_24_absorbers.txt};
\addplot+[line join=bevel,black,mark=square,mark repeat=16,mark phase=6,mark size=2.5] table[x=range_m,y=amplitude_dB] {Figures/measurements/1sphere_time/01_antcal_raw_24_calibrated.txt};
\begin{pgfonlayer}{foreground}
    \node[minimum width=1ex,inner sep=.25ex,circle,fill=white,draw] () at (0.2811,-15.5) {a};
\end{pgfonlayer}

\nextgroupplot[
    xmin=-1,
    xmax=15,
    width=0.625\columnwidth,
    xlabel={\hspace*{-2cm}Normalized range (m)},
    legend entries={Without absorbers,With absorbers,Calibrated},
]
\addplot+[line join=bevel,blue,mark=o,mark repeat=621,mark phase=67] table[x=range_m,y=amplitude_dB] {Figures/measurements/1sphere_time/01_antcal_raw_24_metal.txt};
\addplot+[line join=bevel,red,mark=diamond,mark repeat=621,mark phase=67] table[x=range_m,y=amplitude_dB] {Figures/measurements/1sphere_time/01_antcal_raw_24_absorbers.txt};
\addplot+[line join=bevel,black,mark=square,mark repeat=621,mark phase=67] table[x=range_m,y=amplitude_dB] {Figures/measurements/1sphere_time/01_antcal_raw_24_calibrated.txt};
\begin{pgfonlayer}{foreground}
    \node[minimum width=1ex,inner sep=.25ex,circle,fill=white,draw] () at (0.2811,-15.5) {a};
    \node[minimum width=1ex,inner sep=.25ex,circle,fill=white,draw] () at (1.2741,-28) {b};
    \node[minimum width=1ex,inner sep=.25ex,circle,fill=white,draw] () at (12,-40) {c};
\end{pgfonlayer}
\end{groupplot}
\end{tikzpicture}
    \caption{
        Range-resolved antenna calibration measurement that allows identifying distinct reflections geometrically.
        All values are normalized to line-of-sight distance between Tx and Rx.
        Absorbers reduce reflections~\encircle{a} from the antenna mounting flange (\autoref{fig:mechanic_radial_roll_positioner}\,\encircle{f}) by almost 20\,dB.
        The radial positioners (\autoref{fig:mechanic_radial_roll_positioner}\,\encircle{c}) contribute another strong reflection~\encircle{b}.
        Double reflections~\encircle{c} between gantries are noticeable both with and without absorbers on the antenna flanges (\autoref{fig:native_txrx}\,\encircle{c}).
    }
    \label{fig:antenna_cal_impresp}
\end{figure}

\begin{figure}
	\centering
	\begin{tikzpicture}[>=stealth',paint/.style={draw=#1, fill=#1}]

\draw [line width=1pt] (0,0) circle (0.3);

\draw  [decorate,decoration={shape backgrounds, shape=circle}, paint=blue] [blue, line width=1pt] (-5.6,0) -- ++ (5.6,0.4) arc (90:-90:0.4) -- (190:5.6);
\draw [decorate, decoration={shape backgrounds, shape=diamond}, paint=red] [red, line width=1pt] (-5.6,0) -- ++ (-0.23:5.3) -- (190:5.6);

\node at (-5.6,0) [fill=black, circle, inner sep=2.5pt](RX){};
\node [above = 2pt of RX] {Rx};
\node at (190:5.6) [fill=black, circle, inner sep=2.5pt](TX){};
\node [below = 2pt of TX] {Tx};

\filldraw[fill = black,line width=0.1] ($(RX)+(45:0.1)$) 
                        -- ($(RX)+(20:0.4)$) 
                        -- ($(RX)+(-20:0.4)$) 
                        -- ($(RX)+(-45:0.1)$);
\filldraw[fill = black] ($(TX)+(55:0.1)$) 
                        -- ($(TX)+(30:0.4)$) 
                        -- ($(TX)+(-10:0.4)$) 
                        -- ($(TX)+(-35:0.1)$);

\draw [line width = 1pt] (RX.center) -- ++ (-0.3,0);
\draw [line width = 1pt] (TX.center) -- ++ (10:-0.3);
\draw [<->, line width = 1pt] ($(RX)+(-0.2,0)$) arc (180:190:5.8);
\node at (185:6.6){$\azi-\azo$};
\end{tikzpicture}
    \caption{
        Bistatic path model for the radar reflectivity of a metal sphere comprising reflection at bisector of bistatic azimuth angle and diffraction around the sphere.
        Line styles correspond to \autoref{fig:sphere_impresp}.
    }
    \label{fig:sphere_bistatic_paths}
\end{figure}

\subsection{Wideband Reflectivity Measurement Calibration}
\label{sec:sphere_time_domain}

Antenna calibration and deconvolution is essential for accurate wideband reflectivity measurement.
To exemplify this, we retain the metal sphere from the previous example, but analyze it in time-domain with 16\,GHz bandwidth and thus 1.875\,cm spatial resolution.
The raw measurement results shown in the top plot of \autoref{fig:sphere_impresp} exhibit a diffuse impulse response after the initial reflection from the sphere.
This diffuse response is caused by the impulse response of the measurement antennas in their installed state.

To analyze, calibrate, and subsequently remove the antenna impulse response, we measured the combined response of both antennas as follows:
\Gls{tx} and \gls{rx} were positioned in anti-parallel direction in the horizontal plane without a target present (\autoref{fig:rcs_sphere_setup} without~\encircle{a}).
Beforehand, the \gls{vna} was thru-calibrated directly at the antenna connectors to remove all cabled active and passive components (cf. \autoref{sec:rf_vna}) from the measurement.
Results are illustrated in \autoref{fig:antenna_cal_impresp}.
For comparison, we measured with and without the frontal absorbers shown in \autoref{fig:native_txrx}\,\encircle{c}.
Note the parasitic double reflections (\autoref{fig:antenna_cal_impresp}\,\encircle{c}) that motivate the chosen 30\,m spatial ambiguity (see \autoref{sec:rf_vna}) to enable the removal of such parasitic effects via time-gating.

The wide and shallow impulse response of the installed measurement antennas is inherently convolved with any measured reflectivity.
This necessitates meticulous de-embedding of equipment and propagation effects from measurement data to determine undistorted reflectivity values, which comprise the target only.
Here, we employ the algorithm derived in~\cite{andrich25_eucap_antenna_deconvolution_radar_refl}, to which we refer the interested reader.
\autoref{fig:antenna_cal_impresp} includes the antenna impulse response calibration result used to post-process all subsequent measurements.
See \autoref{fig:sphere_impresp} for the effect of the deconvolution, which improves measurement quality and dynamic range significantly.

Although the calibration was measured with anti-parallel \gls{tx} and \gls{rx}, it is valid for all bistatic reflectivity constellations, as both antennas always point at the target in the center.
Therefore, the calibration is constrained to well within the antenna main lobes and thus limited to medium-sized targets.
Larger targets may require a synthetic aperture approach for antenna calibration, which is beyond the scope of this paper.
However, this example motivates the significance of antenna measurement with respect to \gls{isac}.

\begin{figure*}
    \centering
    \begin{tikzpicture}[spy using outlines={circle, magnification=4, size=2cm, connect spies}]
\pgfdeclarelayer{foreground}
\pgfsetlayers{main,foreground}
\begin{axis}[
        domain=0:360,
        xmajorgrids,
        ymajorgrids,
        grid style={black!15},
        colorbar,
        colormap={WhYlOrRd}{
            rgb=(1.0000, 1.0000, 1.0000),
            rgb=(0.8702, 0.9213, 0.9685),
            rgb=(0.7752, 0.8583, 0.9368),
            rgb=(0.6173, 0.7909, 0.8818),
            rgb=(0.4171, 0.6806, 0.8382),
            rgb=(0.2563, 0.5700, 0.7752),
            rgb=(0.1271, 0.4402, 0.7075),
            rgb=(0.0314, 0.3141, 0.6065),
            rgb=(0.0314, 0.1882, 0.4196)
        },
        colorbar style={
            ylabel={Reflectivity (dBsm)},
            ytick={-50,-45,...,-10},
            yticklabel style={
                text width=1.5em,
                align=right
            }
        },
        width=0.88\linewidth,
        height=8cm,
        enlargelimits=false,
        axis on top,
        tick align=outside,
        tick pos=left,
        xtick pos=bottom,
        xlabel={Turntable angle (deg)},
        ylabel={Normalized range (m)},
        xtick distance=15,
        ytick distance=0.2,
        x tick label style={rotate=90,anchor=east}
    ]
    \addplot[point meta min=-50, point meta max=-10] graphics[xmin=0, xmax=360, ymin=-1.5, ymax=1.5] {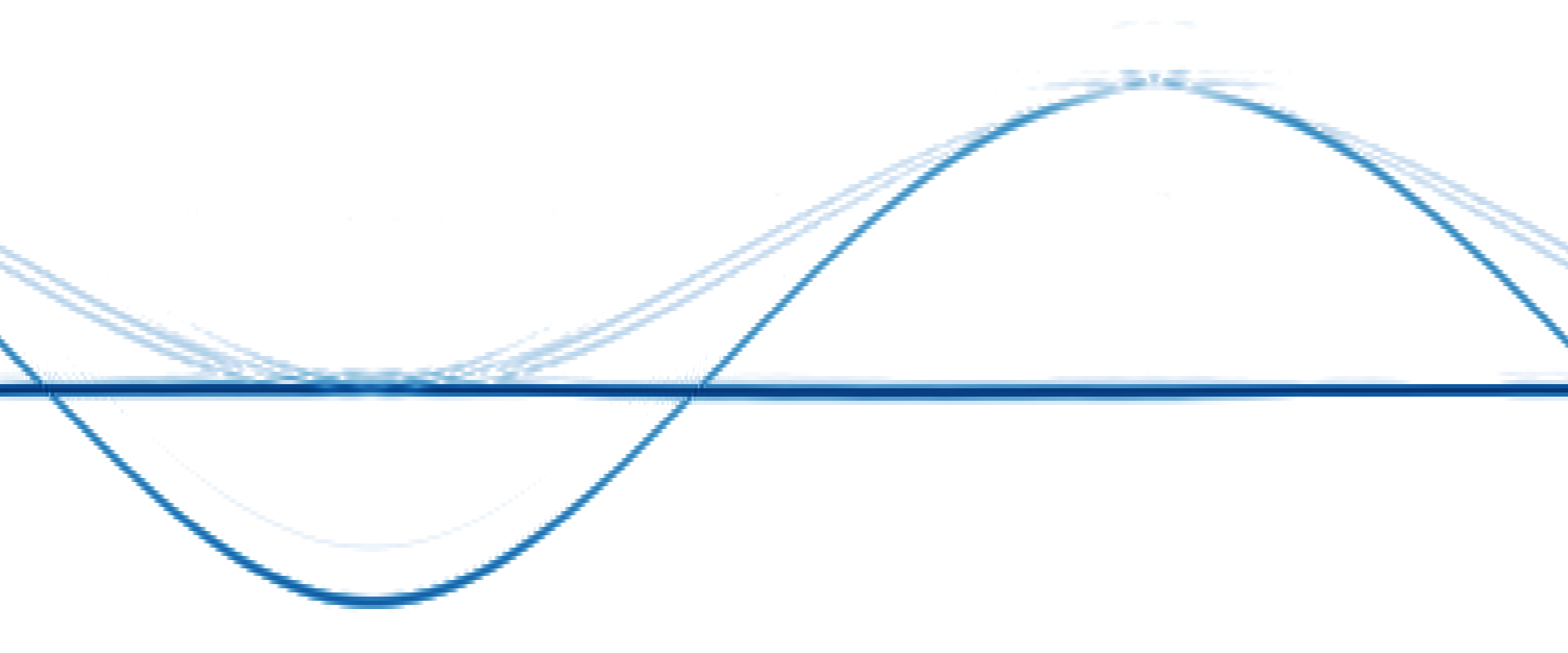};
    \addplot+[mark=diamond, black, mark repeat = 6, mark phase = 6, samples=234, only marks] table[x=ttdeg,y=pathm] {Figures/measurements/2spheres/20_doublesphere_tt_qmo.txt};
    \begin{pgfonlayer}{foreground}
        \node[minimum width=1ex,inner sep=.25ex,circle,fill=white,draw] () at (335,-0.425) {a};
        \node[minimum width=1ex,inner sep=.25ex,circle,fill=white,draw] () at (45,-1.17) {b};
        \node[minimum width=1ex,inner sep=.25ex,circle,fill=white,draw] () at (265,0.81) {c};
        \node[minimum width=1ex,inner sep=.25ex,circle,fill=white,draw] () at (212,-0.90) {d};
        \node[minimum width=1ex,inner sep=.25ex,circle,fill=white,draw] () at (180,0.60) {e};
        \node[minimum width=1ex,inner sep=.25ex,circle,fill=white,draw] () at (85,-0.45) {f};
    \end{pgfonlayer}
    \coordinate (spy1) at (axis cs:160,-0.3);
    \coordinate (spy2) at (axis cs:85,-0.3);
    \coordinate (spy3) at (axis cs:265,1.15);
    \coordinate (mag1) at (axis cs:180,-0.9);
    \coordinate (mag2) at (axis cs:85,0.7);
    \coordinate (mag3) at (axis cs:265,0.3);
\end{axis}
\spy[black, size=2cm] on (spy1) in node[fill=white] at (mag1);
\spy[black, size=3cm, magnification=3] on (spy3) in node[fill=white] at (mag2);
\end{tikzpicture}
    \caption{
        Range-resolved, quasi-monostatic reflectivity measurement of two metal spheres with a large \spherebig\ diameter sphere in the center~\encircle{a} and a smaller \spheresmall\ sphere orbiting on a 0.6\,m radius~\encircle{b}.
        Geometric model for the small sphere reflection \gls{los} path is indicated by diamonds.
        Noteworthy effects include:
        shadowing~\encircle{c} of the small sphere by the large sphere as pictured in \autoref{fig:double_sphere_pic},
        direct interference~\encircle{d} between paths of both spheres,
        secondary reflections~\encircle{e} between both spheres, and
        partial shadowing of large sphere combined with multipath interference~\encircle{f}.
    }
    \label{fig:2sphere_impresp}
\end{figure*}

\begin{figure}
	\centering
    \begin{tikzpicture}
        \node[anchor=south west,inner sep=0] (image) at (0,0) {\includegraphics[width=1\linewidth]{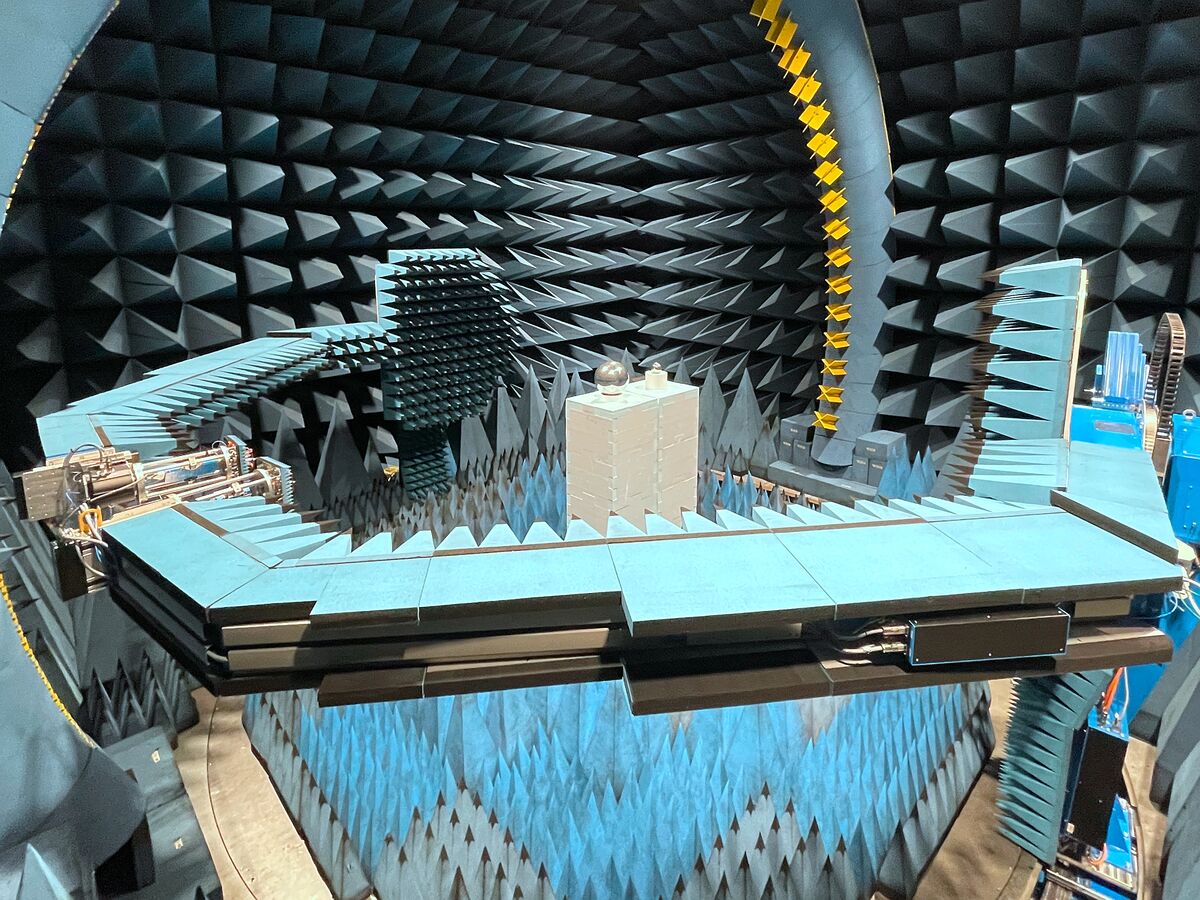}};
        \begin{scope}[x={(image.south east)},y={(image.north west)}]
            \node[minimum width=1ex,inner sep=.25ex,circle,fill=white,draw] () at (0.51,0.52) {a};
            \node[minimum width=1ex,inner sep=.25ex,circle,fill=white,draw] () at (0.5475,0.635) {b};
            \node[minimum width=1ex,inner sep=.25ex,circle,fill=white,draw] () at (0.07,0.37) {c};
            \node[minimum width=1ex,inner sep=.25ex,circle,fill=white,draw] () at (0.04,0.45) {d};
            \node[minimum width=1ex,inner sep=.25ex,circle,fill=white,draw] () at (0.21,0.10) {e};
        \end{scope}
    \end{tikzpicture}
    \caption{
        Quasi-monostatic double sphere measurement setup comprised of central \spherebig\ sphere~\encircle{a} and excentric \spheresmall\ sphere~\encircle{b}.
        Tx~\encircle{c} and Rx~\encircle{d} are stationary.
        The spheres are rotated via the turntable~\encircle{e}, which is depicted at angle 270\textdegree{} of its clockwise machine coordinate system.
    }
    \label{fig:double_sphere_pic}
\end{figure}

\subsection{Spatially Extended Target: Double Sphere}
\label{sec:examples_rcs_double_sphere}

To exemplify a spatially extended target that can still be modeled geometrically, we added an excentric \spheresmall\ diameter sphere orbiting on a radius of 0.6\,m around the larger centered sphere taken over from the previous example.
Both \gls{tx} and \gls{rx} remained stationary at a bistatic angle $\azi - \azo = 10^\circ$ as pictured in \autoref{fig:double_sphere_pic}.
Rotating the turntable by 360\textdegree{}, we observed the reflectivity of an easily modeled extended target.
The calibrated reflectivity is displayed in \autoref{fig:2sphere_impresp}.

Located in the center of the turntable, the reflection from the larger sphere is almost constant, with only minimal excentricity visible from a slight range variation of the peak (\autoref{fig:2sphere_impresp}\,\encircle{a}).
The smaller sphere reflection path matches the geometric \gls{los} model, unless shadowed by the larger sphere, where diffraction causes both interference patterns and prolonged paths (\autoref{fig:2sphere_impresp}\,\encircle{c}).
A similar effect occurs when the smaller sphere partially shadows the larger sphere, also causing interference (\autoref{fig:2sphere_impresp}\,\encircle{f}).
Interaction between both spheres excites additional paths that do not occur when either sphere is measured individually.
Note that the plotted ranges correspond to measured round trip distances.

This straightforward example illustrates how extended targets exhibit complex reflectivity patterns.
Consequently, robust models are required to aid in the classification of bistatic radar returns in the larger context of \gls{isac}.

\begin{figure*}
  \centering
  \begin{minipage}[c]{0.125\linewidth}
    \centering
    \begin{overpic}[width=\linewidth]{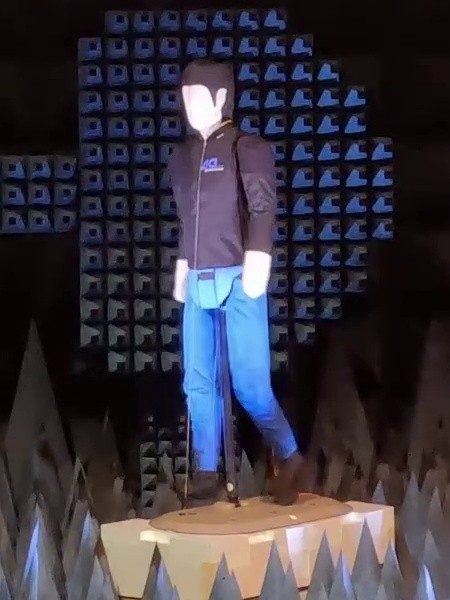}
      \put(42,52){\color{red} \linethickness{.5mm} \vector(20,4){21}}
      \put(28,52){\color{red} \linethickness{.5mm} \vector(-20,-2){21}}
      \put(11,20){\color{red} \linethickness{.5mm} \vector(20,3){21}}
    \end{overpic}
    \footnotesize
        (a) First stepping phase:
        The arms are swinging and the right leg is moving to the back.
  \end{minipage}
  \hfill
  \begin{minipage}[c]{0.35\linewidth}
    \centering
    \vspace{-5px}
\begin{tikzpicture}[>=stealth']
\tikzstyle{every node}=[font=\small]

\definecolor{color0}{rgb}{0.12156862745098,0.466666666666667,0.705882352941177}
\definecolor{color1}{rgb}{1,0.498039215686275,0.0549019607843137}

\begin{axis}[
width=0.72\linewidth,
enlargelimits=false,
tick align=outside,
tick pos=left,
axis on top,
xmin=4, xmax=10, ymin=-4, ymax=4,
xlabel={Range (m)},
ylabel={Doppler speed (m/s)},
xtick style={color=black},
ytick style={color=black},
xmajorgrids,
ymajorgrids,
grid style={black!15},
colorbar,
colormap={bernd}{
    rgb=(1.0                ,  1.0                ,  1.0                )
    rgb=(0.77647058823529413,  0.85882352941176465,  0.93725490196078431)
    rgb=(0.41960784313725491,  0.68235294117647061,  0.83921568627450982)
    rgb=(0.12941176470588237,  0.44313725490196076,  0.70980392156862748)
    rgb=(0.03137254901960784,  0.31764705882352939,  0.61176470588235299)
    rgb=(0.02137254901960784,  0.21764705882352939,  0.41176470588235299)
    rgb=(0.0,  0.0,  0.0)
},
colorbar style={
    ylabel=Norm. power (dB),
    at={(1.1,0)},anchor=south west,
    width=8px
    },
point meta min=-60,
point meta max=0,
]

\addplot graphics [xmin=0, xmax=14.6, ymin=-12.25, ymax=12.5, zmin=-60, zmax=0] {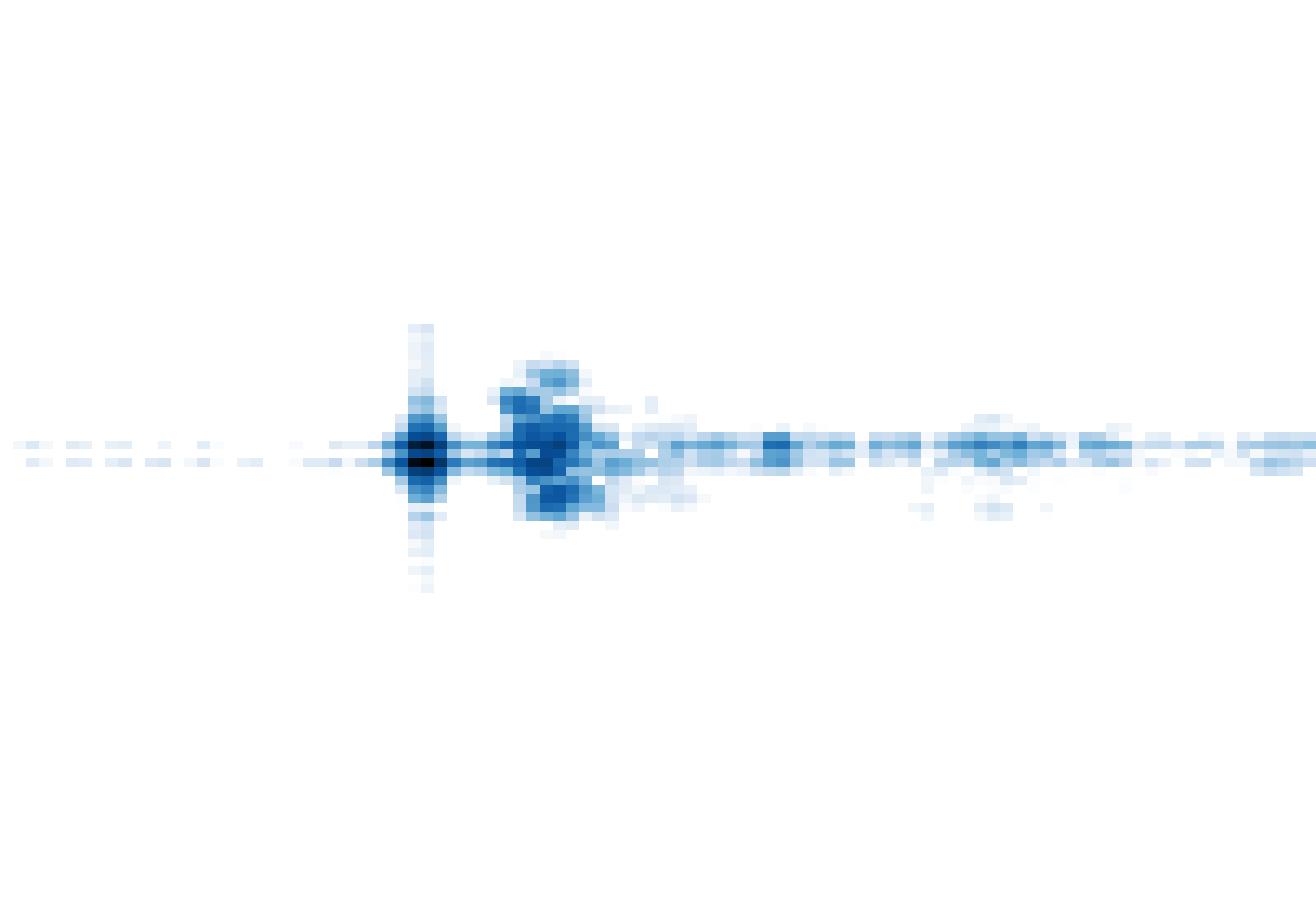};

\end{axis}

\draw[<-, line width= 0.7pt] (1.25,1.7) --++ (.3,0) node[right] {Leg} ;
\draw[<-, line width= 0.7pt] (1.4,0.8) --++ (.3,0) node[right] {Arm} ;
\draw[<-, line width= 0.7pt] (0.65,1.65) --++ (-.15,.15) node[above,] {Arm} ;
\draw[<-, line width= 0.7pt] (0.45,0.8) --++ (.1,-.3) node[below] {LoS} ;
\end{tikzpicture}
\vspace{-5px}
    \vspace{1em}

    \footnotesize
        (b) Micro-Doppler signature:
        All three moving limbs can be individually identified in the ranger-Doppler plot by their peaks.
        The \gls{sdr} system achieves a resolution (size of one pixel) of 15\,cm by 25\,cm\,s\textsuperscript{--1}.
  \end{minipage}
  \hfill
  \begin{minipage}[c]{0.125\linewidth}
    \centering
    \begin{overpic}[width=\linewidth]{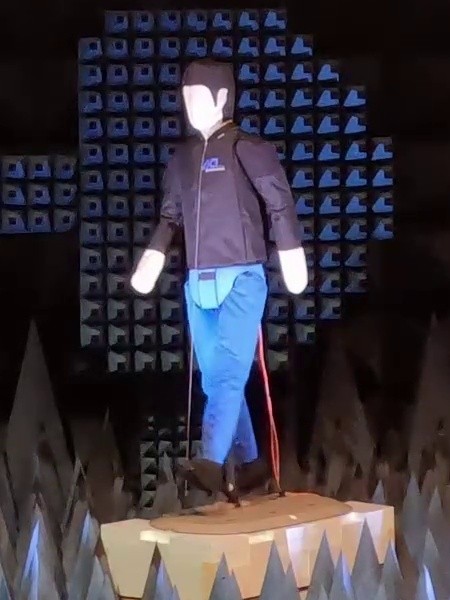}
    \put(28,23){\color{red} \linethickness{.5mm} \vector(-20,7){21}}
    \end{overpic}
    \footnotesize
        (c) Second stepping phase:
        The left leg is kicking forward.
        The other limbs almost reached their destination.
  \end{minipage}
  \hfill
  \begin{minipage}[c]{0.35\linewidth}
    \centering
    \vspace{-5px}
\begin{tikzpicture}[>=stealth']
\tikzstyle{every node}=[font=\small]

\definecolor{color0}{rgb}{0.12156862745098,0.466666666666667,0.705882352941177}
\definecolor{color1}{rgb}{1,0.498039215686275,0.0549019607843137}

\begin{axis}[
width=0.72\linewidth,
enlargelimits=false,
tick align=outside,
tick pos=left,
axis on top,
xmin=4, xmax=10, ymin=-4, ymax=4,
xlabel={Range (m)},
ylabel={Doppler speed (m/s)},
xtick style={color=black},
ytick style={color=black},
xmajorgrids,
ymajorgrids,
grid style={black!15},
colorbar,
colormap={bernd}{
    rgb=(1.0                ,  1.0                ,  1.0                )
    rgb=(0.77647058823529413,  0.85882352941176465,  0.93725490196078431)
    rgb=(0.41960784313725491,  0.68235294117647061,  0.83921568627450982)
    rgb=(0.12941176470588237,  0.44313725490196076,  0.70980392156862748)
    rgb=(0.03137254901960784,  0.31764705882352939,  0.61176470588235299)
    rgb=(0.02137254901960784,  0.21764705882352939,  0.41176470588235299)
    rgb=(0.0,  0.0,  0.0)
},
colorbar style={
    ylabel=Norm. power (dB),
    at={(1.1,0)},anchor=south west,
    width=8px
    },
point meta min=-60,
point meta max=0,
]

\addplot graphics [xmin=0, xmax=14.6, ymin=-12.25, ymax=12.5, zmin=-60, zmax=0] {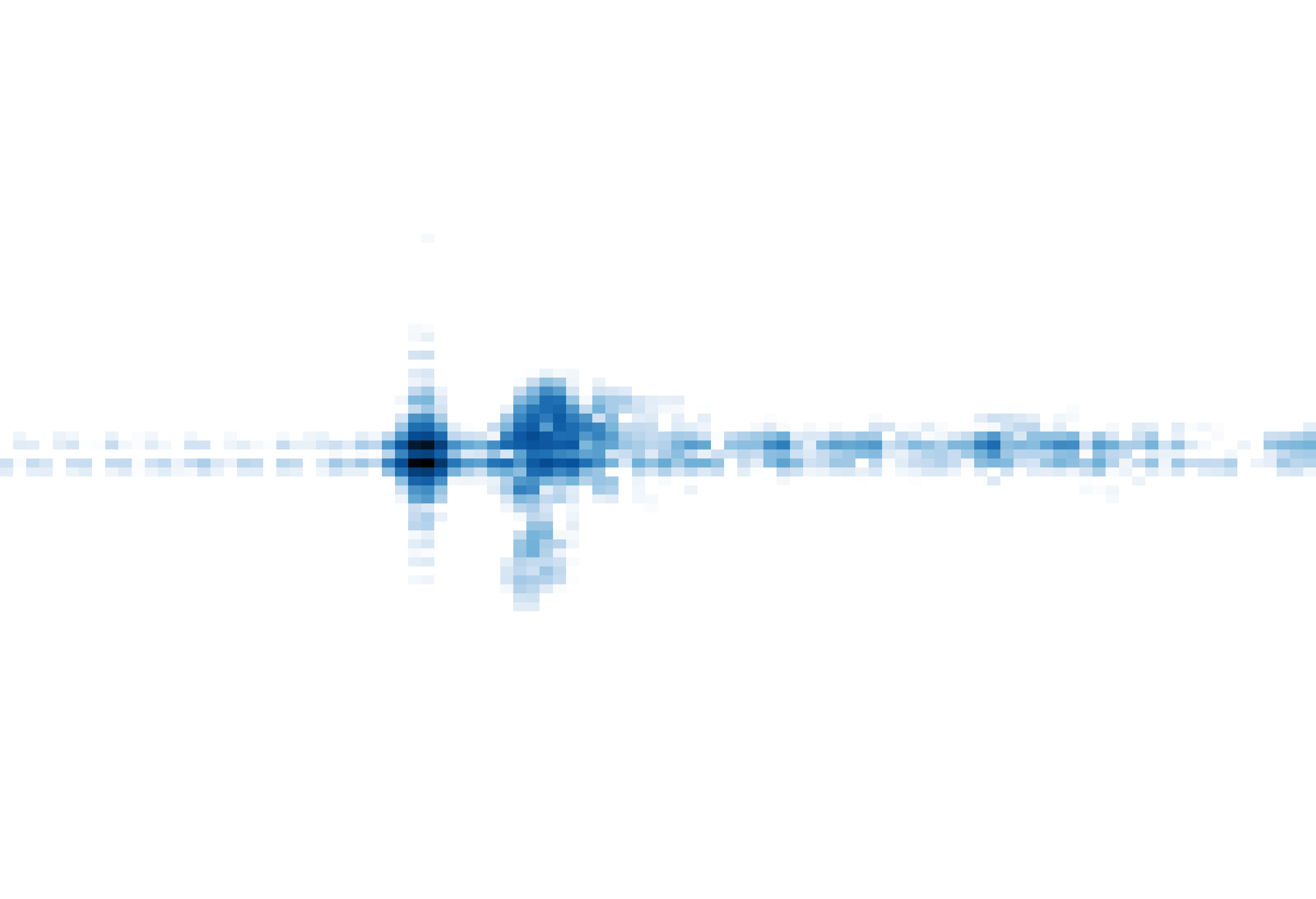};

\end{axis}

\draw[<-, line width= 0.7pt] (1.2,0.35) --++ (.3,0) node[right, align=center] {Kicking\\Leg} ;
\draw[<-, line width= 0.7pt] (0.5,1.45) --++ (.15,.35) node[above] {LoS} ;
\end{tikzpicture}
\vspace{-5px}
    \vspace{1em}

    \footnotesize
        (d) Micro-Doppler signature:
        The forward-kicking leg can be identified by its broad peak dominating the ranger-Doppler plot.
        The \gls{sdr} system achieves a resolution (size of one pixel) of 15\,cm by 25\,cm\,s\textsuperscript{--1}.
    \end{minipage}
    \caption{
        \Gls{isac} scenario:
        A pedestrian dummy is captured by a bistatic micro-Doppler radar setup.
        The subfigures show the dummy states (a, c) and the related ranger-Doppler plots (b, d).
        Subfigures (a) and (b) present an instant in the first half of a step, while (c) and (d) relate to the second half of a step.
        The arrows indicate the moving limbs.
        All limbs can be separated and individually identified in range and Doppler.
        Additionally, the \gls{los} appears, due to the moving gantries.
    }
    \label{fig:uDopplerMeasExample}
\end{figure*}

\subsection{Wideband Micro-Doppler Signature of Pedestrian}
\label{sec:examples_micro_doppler}

\begin{figure}
  \centering
    \includegraphics[width=\linewidth]{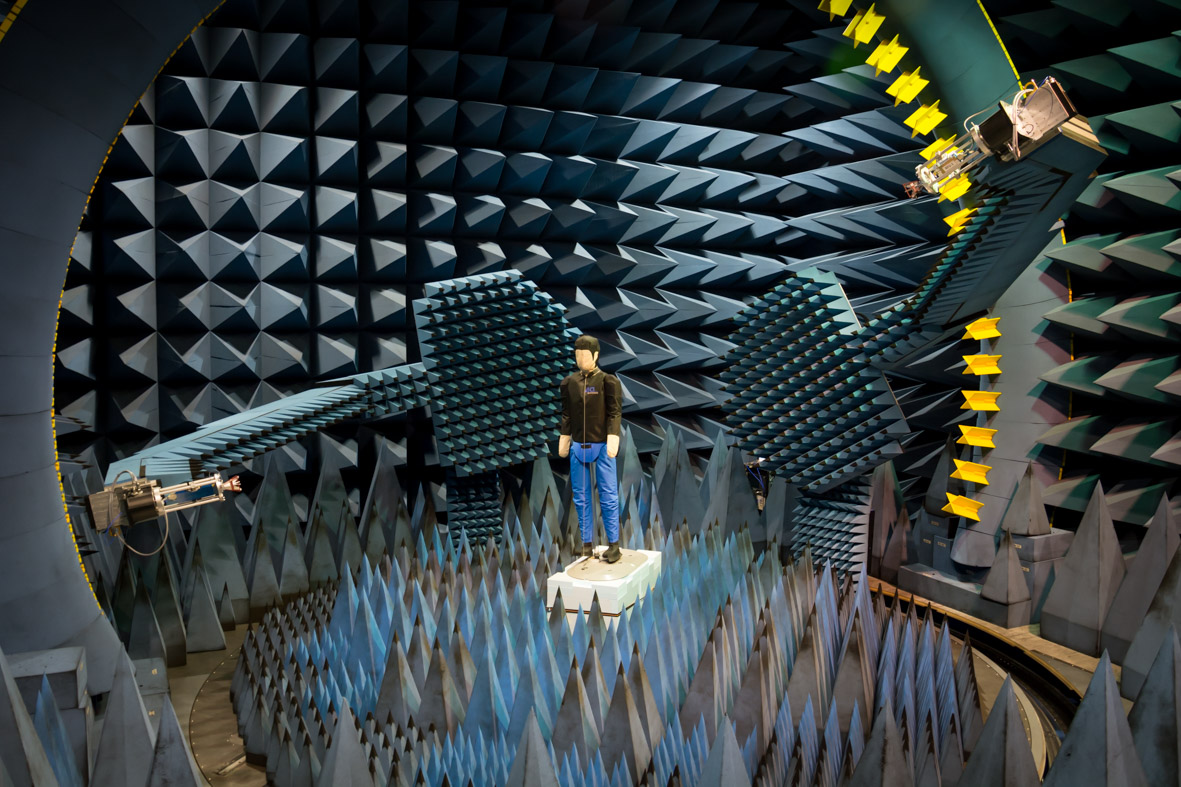}
    \caption{
        Bistatic micro-Doppler measurement setup:
        A motorized pedestrian dummy is illuminated and observed by two distributed, wideband \gls{sdr} transceivers, mounted on the \gls{bira} gantries.
    }
    \label{fig:uDopplerTotale}
\end{figure}

Whereas the previous examples relied on the static measurement of metal spheres as canonical reference objects, our next bistatic example for \gls{bira} opts for a more realistic application:
The micro-Doppler measurement of a moving pedestrian \gls{ncap} dummy.
We employed our wideband \gls{sdr} system as distributed \gls{tx} and \gls{rx} as described in \autoref{sec:rf_rfsoc}.
A motorized \gls{vru} pedestrian dummy served as a dynamic target that moves its limbs to mimic a walking motion, imprinting its micro-Doppler signature on the observed signal.
The measurement setup is depicted in \autoref{fig:uDopplerTotale}.

The \glspl{sdr} rely on Marki MMIQ-0520 IQ converters to transmit and receive a periodic \gls{ofdm} illumination signal between 11\,GHz and 13\,GHz, representing a hypothetical 6G band in the FR3 frequency range.
The observed signal was continuously recorded at a data rate of 10\,GB/s for subsequent offline processing, e.g., using \gls{ai}-supported algorithms.
Simultaneously, the \gls{tx} moved in both azimuth and co-elevation to emulate a non-stationary illuminator and to capture the micro-Doppler signature of the dummy from a wide range of perspectives, i.e., bistatic angle combinations.
Two measurement snapshots are illustrated in \autoref{fig:uDopplerMeasExample}.

\autoref{fig:uDopplerMeasExample}a and \autoref{fig:uDopplerMeasExample}c show the dummy in two phases of a step.
The corresponding frames of the continuously generated ranger-Doppler plots are presented in \autoref{fig:uDopplerMeasExample}b and \autoref{fig:uDopplerMeasExample}d.
Due to the moving gantries, the \gls{los} is evident in both ranger-Doppler plots.
First, in \autoref{fig:uDopplerMeasExample}a, one leg and arm are seen swinging back, while the other arm was moving forward.
All three limbs can clearly be identified as distinct peaks in range and Doppler in \autoref{fig:uDopplerMeasExample}b.
An instant later, cf. \autoref{fig:uDopplerMeasExample}c, the other foot performed a kick-like forward motion.
While the other limbs are still slightly visible, the kicking leg dominates the lower half of the ranger-Doppler plot in \autoref{fig:uDopplerMeasExample}d.
Compared to the first phase, the higher speed of this inner movement can easily be noticed.
Considering one cycle of the movement, the peaks of all limbs move around the center on elliptical curve sectors.
This leads to the time-dependent micro-Doppler signature of the target that changes between bistatic configurations.

The 2\,GHz instantaneous measurement bandwidth of our \gls{sdr} measurement system enables a high range resolution of 15\,cm.
Comparable bandwidths and therefore resolutions are to be expected with \gls{isac} and 6G, enabling simultaneously resolving inner target motion in both range- and Doppler-domain.
The present example illustrates this by individually resolving the limbs of the pedestrian dummy.
Similarly, other target objects are identifiable via their unique micro-Doppler signature, e.g., \gls{uas}~\cite{costa24_arxiv_uav_rcs_udoppler, costa24_radarconf_uav_udoppler_model, costa25_arxiv_vtol, costa25_arxiv_model}.

Equipped with its wideband \gls{sdr} transceiver capability, \gls{bira} is uniquely situated to gather micro-Doppler signatures for a wide variety of target objects from small \gls{uas} up to passenger cars.
These signatures enable, e.g., training of \gls{ai} models for target detection and classification.

\begin{table*}
\centering
\caption{Parameters of comparative measurements of a Schwarzbeck CTIA\,0710 antenna.}
\label{table:antenna_measurements}
\footnotesize
\renewcommand{\arraystretch}{1.1}
\begin{tabular}{ |l|p{4.2cm}|p{4.2cm}|p{4.2cm}| }
\hline
 & \multicolumn{1}{c|}{\textbf{BIRA}} & \multicolumn{1}{c|}{\textbf{MVG SG~3000F}} & \multicolumn{1}{c|}{\textbf{NSI-800F-10x}} \\
 & using static gantry with \gls{rx} probe, see \autoref{fig:antenna_measurment_setup} \encircle{d}+\encircle{e} & Microwave Vision Group arch, \newline see \autoref{fig:antenna_measurment_setup} \encircle{c} & Antenna measurement laboratory \newline w/ NSI-800F-10x and NSI-SW5305 \\ \hline
 target distance & 3\,m & 4~m  & 5.85~m \\ \hline
 azimuth grid & 3° step size (timining constraints) & 3° step size (comparable to \gls{bira}) & 3° step size (comparable to \gls{bira}) \\ \hline
 elevation grid & 1° step size (comparable to MVG) & 1° (arch probe antenna spacing) & 1° step size (comparable to MVG) \\ \hline
 frequency grid & 2\,...\,8 GHz (100 MHz steps) & 2\,...\,6 GHz (100 MHz steps) & 2\,...\,8 GHz (100 MHz steps) \\ \hline
 approximate duration & 16\,h & 1.5\,h & 35~h \\ \hline
\end{tabular}
\end{table*}

\subsection{Antenna Directivity Pattern}
\label{sec:examples_antenna}

Having used \gls{bira} already for antenna calibration in terms of their impulse response within main lobe, see \autoref{sec:sphere_time_domain}, we finally validate the use case of antenna beam pattern measurements.
Here, the most important benefit of \gls{bira} is the frequency coverage of up to 260~GHz (see \autoref{table:rf}) and the possibility  of antenna measurements in their installed state on large targets, e.g., cars.
As a proof of principle, we carried out comparative measurements with \gls{bira}, the \gls{bira} antenna measurement arch, and our on-campus \glsfirst{aml}~\cite{tayyab25_ursi_rsl_5g_antenna_automotive}, see \autoref{table:antenna_measurements}.

\begin{figure}
	\centering
    \begin{tikzpicture}
        \node[anchor=south west,inner sep=0] (image) at (0,0) {\includegraphics[width=1\linewidth]{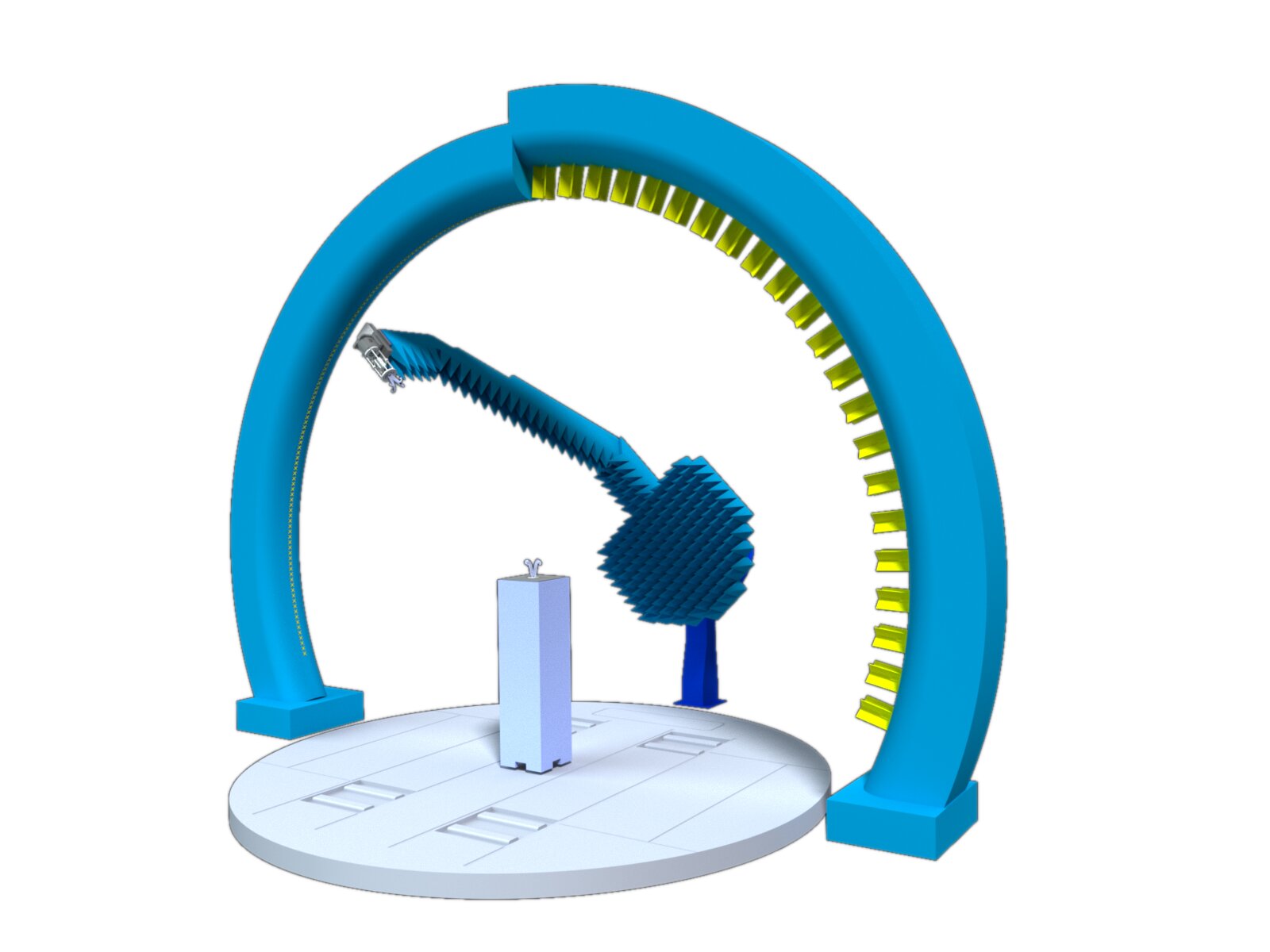}};
        \begin{scope}[x={(image.south east)},y={(image.north west)}]
            \node[minimum width=1ex,inner sep=.25ex,circle,fill=white,draw] () at (0.42,0.35) {a};
            \node[minimum width=1ex,inner sep=.25ex,circle,fill=white,draw] () at (0.50,0.15) {b};
            \node[minimum width=1ex,inner sep=.25ex,circle,fill=white,draw] () at (0.20,0.60) {c};
            \node[minimum width=1ex,inner sep=.25ex,circle,fill=white,draw] () at (0.53,0.42) {d};
            \node[minimum width=1ex,inner sep=.25ex,circle,fill=white,draw] () at (0.33,0.57) {e};
        \end{scope}
    \end{tikzpicture}
    \caption{Antenna measurement with \gls{bira} vs.\ \gls{mvg} arch:
        Styrodur base with the \gls{aut}~\encircle{a} on the turntable~\encircle{b}.
        \Gls{mvg} measurement arch~\encircle{c} (frequency range 0.4\ldots 6~GHz). 
        \Gls{bira} static gantry~\encircle{d} with probe antenna~\encircle{e} (frequency ranges: see \autoref{table:rf}).
        Note: During the \gls{mvg} measurements, the moving \gls{bira} gantry was absent.}
    \label{fig:antenna_measurment_setup}
\end{figure}

\Glsfirst{aut} was a Schwarzbeck CTIA\,0710, mounted in the phase center.
For \gls{bira}, the same antenna model was used as probe, mounted on the static gantry~(\autoref{fig:antenna_measurment_setup}).
All three systems followed the spherical near-field principle~\cite{hansen_1988}, whereby far-field conditions were fulfilled for the \gls{aut} in the investigated frequency range.

\begin{figure}
    \centering
    \begin{tikzpicture}
\begin{polaraxis}[
   width=\columnwidth,
   xticklabel=$\pgfmathprintnumber{\tick}^\circ$,
   xtick={-90,-60,...,90},
   ytick={-30,-25,...,0},
   xmin=-90,xmax=90,
   ymin=-30, ymax=0,
   y coord trafo/.code=\pgfmathparse{#1+30},
   y coord inv trafo/.code=\pgfmathparse{#1-30},
   rotate=90,
   xticklabel style={anchor=\tick-90},
   xticklabels={90$^\circ$,60$^\circ$,30$^\circ$,0$^\circ$,30$^\circ$,60$^\circ$,90$^\circ$},
   yticklabel style={anchor=north, xshift=0cm,yshift=0cm},
   yticklabel={\pgfmathprintnumber{\tick} dB},
   yticklabels={--30 dB,,--20,,--10,,0 dB},
    legend style={
  at={(current bounding box.south-|current axis.south)},
  yshift=-0.5cm,
  anchor=north,
  legend columns=-1
},
   legend entries={BIRA\hspace*{1em},MVG \hspace*{1em},AML\hspace*{1em}},]
\addplot+[line join=bevel,mark=none,thick,black,solid] table[x=angle,y=BiRa_NF] {Figures/measurements/antenna/antenna_pattern_CTIA0710.txt};
\addplot+[line join=bevel,mark=none,thick,blue,densely dashdotdotted] table[x=angle, y expr={\thisrow{MVG_FF}-3.051}] {Figures/measurements/antenna/antenna_pattern_CTIA0710.txt};
\addplot+[line join=bevel,mark=none,thick,red,densely dashdotted] table[x=angle,y=AML_FF] {Figures/measurements/antenna/antenna_pattern_CTIA0710.txt};
\end{polaraxis}
\end{tikzpicture}
    \caption{
        Normalized gain of a Schwarzbeck CTIA\,0710 measured with \gls{bira}, \gls{mvg} arch, and \gls{aml}.
        A vertical cut at azimuth angle $\az$~=~0° at f~=~6~GHz is shown.
        The results of the \gls{bira} system show a maximum deviation of 0.16~dB to the \gls{mvg} system and 0.17~dB to the \gls{aml}.
    }
    \label{fig:ant_comp_cuts_normalized}
\end{figure}

\begin{figure}
    \raggedleft
    \begin{tikzpicture}[font=\scriptsize]
\begin{axis}[
   width=\columnwidth,height=4.5cm,
   title=maximum gain,
   every axis title/.style={below right,at={(0,1)},draw=black,fill=white},
   ylabel={$G_{\text{max}(\el,\az)}$ (dBi)},
   xmin=2, xmax=8,
   ymin=7, ymax=13,
   ytick={7,8,...,13},
   grid=major,
   legend style={
  at={(current bounding box.south-|current axis.south)},
  anchor=north,
  legend columns=-1
},
   legend entries={BIRA\hspace*{1em},MVG\hspace*{1em},AML\hspace*{1em},data sheet\hspace*{1em}},
]
\addplot+[line join=bevel,mark=none,thick,black,solid] table[x=f,y=BiRa] {Figures/measurements/antenna/max_gain_CTIA0710.txt};
\addplot+[line join=bevel,mark=none,thick,blue,densely dashdotdotted] table[x=f,y=MVG] {Figures/measurements/antenna/max_gain_CTIA0710.txt};
\addplot+[line join=bevel,mark=none,thick,red,densely dashdotted] table[x=f,y=AML] {Figures/measurements/antenna/max_gain_CTIA0710.txt};
\addplot+[line join=bevel,mark=none,thick,green,densely dotted] table[x=f,y=Datasheet] {Figures/measurements/antenna/max_gain_CTIA0710.txt};
\end{axis}
\end{tikzpicture}

\begin{tikzpicture}[font=\scriptsize]
\begin{axis}[
   width=\columnwidth,height=4.5cm,
   title={maximum gain difference (data sheet as reference)},
   every axis title/.style={below right,at={(0,1)},draw=black,fill=white},
   xlabel={Frequency (GHz)},
   ylabel={$\Delta G_{\text{max}(\el,\az)}$ (dB)},
   xmin=2, xmax=8,
   ymin=-1, ymax=1,
   grid=major,
]
\addplot+[line join=bevel,mark=none,thick,green,densely dotted] table[x=f,y expr={\thisrow{Datasheet}-\thisrow{Datasheet}}] {Figures/measurements/antenna/max_gain_CTIA0710.txt};
\addplot+[line join=bevel,mark=none,thick,black,solid] table[x=f,y expr={\thisrow{Datasheet}-\thisrow{BiRa}}] {Figures/measurements/antenna/max_gain_CTIA0710.txt};
\addplot+[line join=bevel,mark=none,thick,blue,densely dashdotdotted] table[x=f,y expr={\thisrow{Datasheet}-\thisrow{MVG}}] {Figures/measurements/antenna/max_gain_CTIA0710.txt};
\addplot+[line join=bevel,mark=none,thick,red,densely dashdotted] table[x=f,y expr={\thisrow{Datasheet}-\thisrow{AML}}] {Figures/measurements/antenna/max_gain_CTIA0710.txt};
\end{axis}
\end{tikzpicture}
    \caption{
        Maximum gain $G_\text{max}(\el,\az$) over frequency of Schwarzbeck CTIA\,0710 antenna, measured with \gls{bira}, \gls{mvg} arch,  and \gls{aml}, compared to the antenna data sheet.
        The frequency range of the \gls{mvg} system  is technically limited to 6~GHz.
    }
    \label{fig:antenna_measurment_compare_gain}
\end{figure}

\autoref{fig:ant_comp_cuts_normalized} shows a vertical cut at azimuth angle $\az$~=~0° at 6~GHz.
The normalized measured values differ by a maximum of 0.17~dB.
Comparing the maximum gain (\autoref{fig:antenna_measurment_compare_gain}) with the data sheet of the antenna, deviations of up to 0.8~dB were found, which is well within the range of typical antenna gain measurement uncertainties of $\pm$1.5~dB.
These exemplary results verify \gls{bira} also as versatile and reliable antenna measurement system that extends the capabilities of \gls{vista} beyond the constraints of the SG~3000F measurement arch, namely a fixed probe distance and a 6\,GHz upper frequency limit.

\subsection{Further Measurements}

The previous examples were chosen as being straightforward, descriptive, and self-contained with the intention to motivate, introduce, and prove the functionality and capabilites of \gls{bira} as well as to illustrate the challenges and opportunities of wideband bistatic radar reflectivity measurement.
More complex examples are beyond the scope of this paper.
However, since its inauguration, \gls{bira} has already facilitated a multitude of productive measurements to which we direct the interested reader.
These include various \gls{uas}~\cite{costa24_arxiv_uav_rcs_udoppler, costa24_radarconf_uav_udoppler_model, costa25_arxiv_vtol, costa25_arxiv_model} and \gls{vru}~\cite{varga25_eucap_antenna_offsets} measurements as well as the characterization of reflective intelligent surfaces~\cite{reher25_gemic_ris}.

\section{Summary}
\label{sec:conclusions}

In this paper, we introduced our \glsfirst{bira} measurement system, which extends the \glsfirst{vista}~\cite{hein23_roe_vista} with two universal mechanical positioners.
Together with target rotation by the turntable, independent illumination and observation angles of both more than a hemisphere (0 to 360\textdegree{} azimuth, 0 to 114\textdegree{} co-elevation) are realized with sub-millimeter accuracy.
This comprehensive upgrade was inaugurated in 2023 and addresses the requirements of research for 6G and beyond in the upcoming decades~\cite{shatov24_access_jrc,rosemann24jcas_JCAS6G,ITU2022,ITU2023}.
Although motivated by and named after bistatic radar, i.e., \gls{isac}, our \gls{bira} system is entirely use-case agnostic and includes a variety of features useful for 3-dimensional antenna pattern measurements up to the sub-THz frequency range.

\Gls{bira} is a modular spherical positioning system for radar targets up to the size of a passenger car.
Either positioner can be installed and used independently and reversibly.
A universal probe flange, power supply, integrated microwave cabling and \gls{lo} distribution, and generic low-level \gls{api} support almost arbitrary payloads and ensure future upgradability.
A digital twin ensures the safe operation of the positioners, facilitates measurement planning for researchers, and accelerates software development.

Currently available microwave probes include coaxial frequency converters for parallel, dual-polarized measurements up to 50\,GHz and linearly polarized waveguide converters for almost continuous signal coverage up to 260\,GHz.
All probes are baseband agnostic and compatible with an integrated \gls{vna} as well as distributed \gls{sdr} transceivers.

\Gls{bira} is a bistatic reflectivity measurement facility suitable for extended \gls{isac} research.
It is one of only two installations with at least four spherical degrees of freedom, which is the minimum required for fully bistatic reflectivity measurements.
Secondly and more importantly, \gls{bira} employs distributed \gls{sdr} transceivers, while other installations use \glspl{vna}~\cite{sota_biancha_2010,sota_biancha_2016,biancha_www,sota_cactus,sota_babi,sota_emsl_sieber_1992,emsl_www,sota_lamp_tian_2021,lamp_www} with theoretically zero instantaneous bandwidth, restricting their applicability to the reflectivity of stationary objects.
In contrast, the instantaneous bandwidth of the \gls{sdr} tranceivers (up to 4\,GHz) enables dynamic \gls{isac} measurements with superior range resolution of up to 3.75\,cm.
\Gls{bira} provides novel and highly relevant experimental access to the bistatic radar reflectivity of extended objects and antenna characterization up to the sub-THz range.

The \gls{thimo} has been conducting leading research of wireless transmission related to road and rail traffic and drones, based on its automotive antenna facility VISTA since 2014~\cite{hein23_roe_vista}.
The \gls{bira} upgrade presents a unique asset of the \gls{thimo} for \gls{isac} research in the context of 6G and beyond.

\section*{Acknowledgments}

We gratefully acknowledge contributions from Michael Huhn, Mario Lorenz, Masoumeh Pourjafarian, Marius Schmidt, and Isabella Varga supporting the operation and improvement of BIRA.
We also gratefully acknowledge the contributions from Thomas Dallmann and Dirk Heberling to the measurement concept and relevant applications.

\IEEEtriggeratref{36}
\bibliographystyle{IEEEtran}
\bibliography{References}

\end{document}